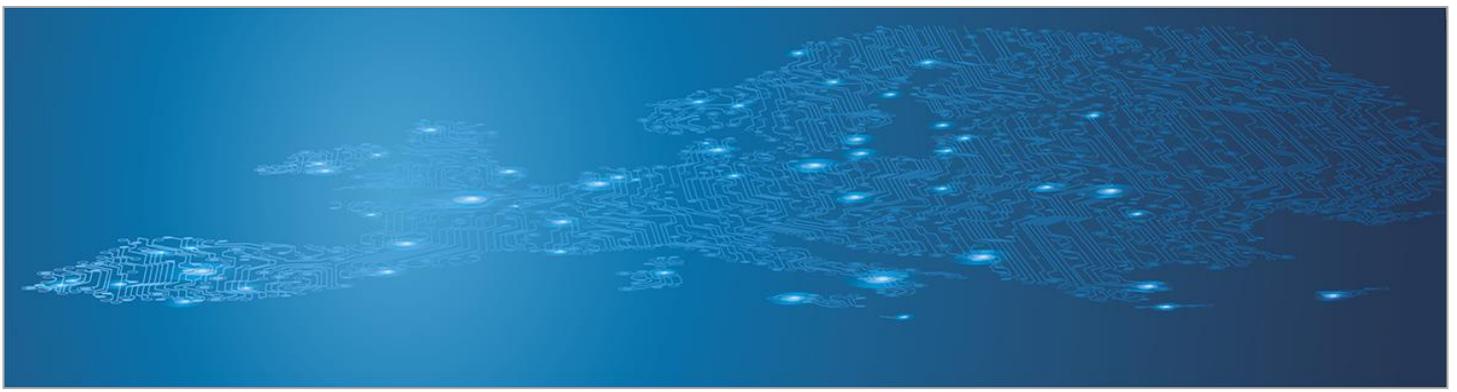

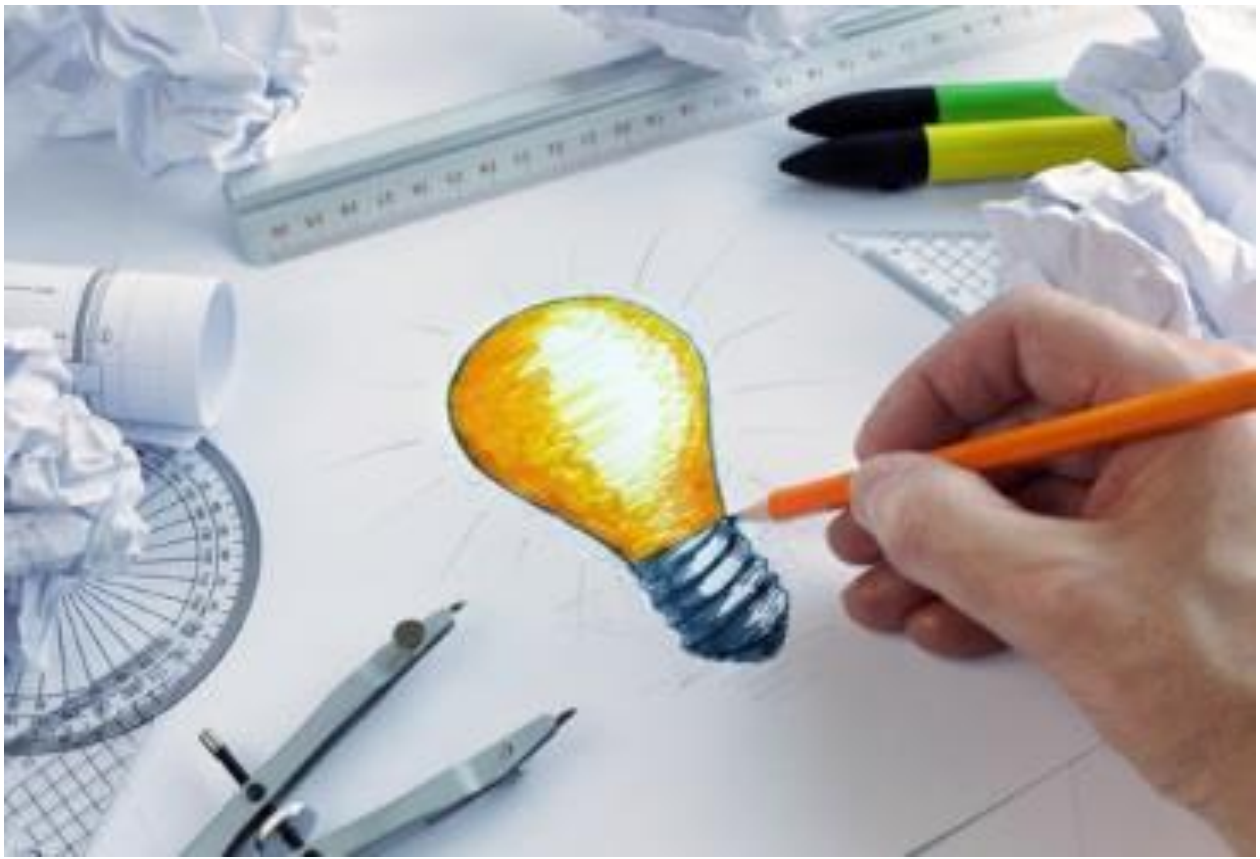

# *Privacy and Data Protection by Design – from policy to engineering*

December 2014

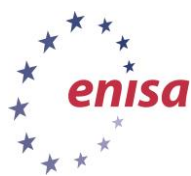

## About ENISA

The European Union Agency for Network and Information Security (ENISA) is a centre of network and information security expertise for the EU, its member states, the private sector and Europe's citizens. ENISA works with these groups to develop advice and recommendations on good practice in information security. It assists EU member states in implementing relevant EU legislation and works to improve the resilience of Europe's critical information infrastructure and networks. ENISA seeks to enhance existing expertise in EU member states by supporting the development of cross-border communities committed to improving network and information security throughout the EU. More information about ENISA and its work can be found at www.enisa.europa.eu.

## Authors

George Danezis, Josep Domingo-Ferrer, Marit Hansen, Jaap-Henk Hoepman, Daniel Le Métayer, Rodica Tirtea, Stefan Schiffner

## Contact

For contacting the authors please use sta@enisa.europa.eu

For media enquires about this paper, please use press@enisa.europa.eu.

## Acknowledgements

We have discussed this work with numerous people at various occasions; we thank for the valuable input we got from these discussions. While we cannot name all, we would like to thank Sébastien Gambs and Rosa Barcelo for their feedback on drafts of this report.

# Executive summary

Privacy and data protection constitute core values of individuals and of democratic societies. This has been acknowledged by the European Convention on Human Rights[1] and the Universal Declaration of Human Rights[2] that enshrine privacy as a fundamental right. With the progress in the field of information and communication technologies, and especially due to the decrease in calculation and storage costs, new challenges to privacy and data protection have emerged. There have been decades of debate on how those values—and legal obligations—can be embedded into systems, preferably from the very beginning of the design process.

One important element in this endeavour are technical mechanisms, most prominently so-called Privacy-Enhancing Technologies (PETs), e.g. encryption, protocols for anonymous communications, attribute based credentials and private search of databases. Their effectiveness has been demonstrated by researchers and in pilot implementations. However, apart from a few exceptions, e.g., encryption became widely used, PETs have not become a standard and widely used component in system design. Furthermore, for unfolding their full benefit for privacy and data protection, PETs need to be rooted in a data governance strategy to be applied in practice. The term "Privacy by Design", or its variation "Data Protection by Design", has been coined as a development method for privacy-friendly systems and services, thereby going beyond mere technical solutions and addressing organisational procedures and business models as well. Although the concept has found its way into legislation as the proposed European General Data Protection Regulation[3], its concrete implementation remains unclear at the present moment.

This report contributes to bridging the gap between the legal framework and the available technological implementation measures by providing an inventory of existing approaches, privacy design strategies, and technical building blocks of various degrees of maturity from research and development. Starting from the privacy principles of the legislation, important elements are presented as a first step towards a design process for privacy-friendly systems and services. The report sketches a method to map legal obligations to design strategies, which allow the system designer to select appropriate techniques for implementing the identified privacy requirements. Furthermore, the report reflects limitations of the approach. It distinguishes inherent constraints from those which are induced by the current state of the art. It concludes with recommendations on how to overcome and mitigate these limits.

## Objectives of the report

This report shall promote the discussion on how privacy by design can be implemented with the help of engineering methods. It provides a basis for better understanding of the current state of the art concerning privacy by design with a focus on the technological side. Data protection authorities can use the report as a reference of currently available technologies and methods. Lastly, the report should help regulators to better understand the opportunities, challenges and limits of the by-design principles with respect to privacy and data protection, to improve the expressiveness and effectiveness of future policy.

---

[1] http://conventions.coe.int/Treaty/Commun/QueVoulezVous.asp?NT=005&CM=7&DF=11/12/2014&CL=ENG

[2] http://www.un.org/en/documents/udhr/

[3] http://ec.europa.eu/justice/data-protection/index_en.htm

## Key findings

We observed that privacy and data protection features are, on the whole, ignored by traditional engineering approaches when implementing the desired functionality. This ignorance is caused and supported by limitations of awareness and understanding of developers and data controllers as well as lacking tools to realise privacy by design. While the research community is very active and growing, and constantly improving existing and contributing further building blocks, it is only loosely interlinked with practice. This gap has to be bridged to achieve successful privacy-friendly design of systems and services and evolve the present state of the art. Further, enforcement of compliance with the regulatory privacy and data protection framework has to become more effective, i.e., better incentives for compliance as well as serious sanctions for non-compliance are needed. Also, privacy-by-design can very much be promoted by suitable standards that should incorporate privacy and data protection features as a general rule.

## List of Recommendations

This is a mere list of our main recommendations; for deeper explanations see section 5.2.

- Policy makers need to support the development of new incentive mechanisms for privacy-friendly services and need to promote them.
- The research community needs to further investigate in privacy engineering, especially with a multidisciplinary approach. This process should be supported by research funding agencies. The results of research need to be promoted by policy makers and media.
- Providers of software development tools and the research community need to offer tools that enable the intuitive implementation of privacy properties.
- Especially in publicly co-founded infrastructure projects, privacy-supporting components, such as key servers and anonymising relays, should be included.
- Data protection authorities should play an important role providing independent guidance and assessing modules and tools for privacy engineering.
- Legislators need to promote privacy and data protection in their norms.
- Standardisation bodies need to include privacy considerations in the standardisation process.
- Standards for interoperability of privacy features should be provided by standardization bodies.

# Table of Contents

# 1   Introduction

Privacy is a fundamental human right. This is acknowledged by Article 8 of the European Convention on Human Rights[4], which provides a right to respect for one's "private and family life, his home and his correspondence". Similarly, the Charter of Fundamental Rights of the European Union defines the "respect for private and family life" (Article 7) and adds a specific article on "protection of personal data" (Article 8). Moreover on an even wider scope, Article 12 of the Universal Declaration of Human Rights[5] protects an individual from "arbitrary interference with his privacy, family, home or correspondence," and "attacks upon his honour and reputation". Privacy protection is not only to be regarded as an individual value, but also as an essential element in the functioning of democratic societies.

In the digital world, we observe a massive power imbalance between data processing entities, which determine what and how data is processed, and the individuals whose data is at stake, i.e., whose lives might be influenced by decisions based on automated data analysis, or by failures to adequately protect private information. Hence, specifically in the digital world, the protection of privacy plays a crucial role. However, when using a specific service, many individuals are often unaware of the data processing and its consequences.[6] Moreover, the user's subsequent control over the nature of the processing that happens to their personal data once it is given away is limited. Lastly, penalties for infringements of legal data protection obligations usually take effect only after the fact, i.e. if a breach or misuse of data has occurred already.

At the same time, our society more and more depends on the trustworthy functioning of information and communication technologies (ICT). Processing capabilities—storage, CPUs, networking—have increased, processing of personal data has intensified. Current trends comprise minimisation of processing power of sensors, detachment of guaranteed locations in cloud computing, mobile and "always on" usage, big data analytics for all conceivable purposes. In essence, almost all areas of life are connected with ICT support, and hence would be affected if trustworthiness cannot be maintained. On the down side of this development are unclear responsibilities and lack of transparency for users and regulatory bodies, and largely missing guarantees of privacy and security features. In practice, European Data Protection Authorities lack the capacity to effectively and systematically monitor data processing or penalise premeditated or negligent wrongdoing.

Having said that, digital technologies may be designed to protect privacy. Since the 1980s technologies with embedded privacy features have been proposed [57]. The term "privacy-enhancing technologies (PETs)" was introduced for a category of technologies that minimise the processing of personal data [119, 39]. By using PETs, the risks for the users' informational privacy would decrease and the legal data protection obligations of the entities responsible for the data processing would be fulfilled more easily. In this spirit, the European Commission issued a Communication to promote PETs in 2007 [67], and privacy-enhancing technologies have become a field of their own within computer science, computer security and cryptography, but also of law, social sciences or economics. However, the mere existence of PET concepts or implementations has been proven insufficient to extensively address the challenge of supporting the individual's right to privacy; privacy cannot be guaranteed by just technology, let alone by a few PET components embedded in a bigger ICT system, cf. [148].

---

[4] http://conventions.coe.int/Treaty/Commun/QueVoulezVous.asp?NT=005&CM=7&DF=11/12/2014&CL=ENG
[5] http://www.un.org/en/documents/udhr/
[6] See results of the FP7 project CONSENT http://cordis.europa.eu/result/rcn/140471_en.html

Hence, privacy needs to be considered from the very beginning of system development. For this reason, Cavoukian[7] coined the term "Privacy by Design", that is, privacy should be taken into account throughout the entire engineering process from the earliest design stages to the operation of the productive system. This holistic approach is promising, but it does not come with mechanisms to integrate privacy in the development processes of a system. The privacy-by-design approach, i.e. that data protection safeguards should be built into products and services from the earliest stage of development, has been addressed by the European Commission in their proposal for a General Data Protection Regulation [159]. This proposal uses the terms "privacy by design" and "data protection by design" synonymously. Article 23 ("Data protection by design and by default") would oblige those entities responsible for processing of personal data to implement appropriate technical and organisational measures and procedures both at the time of the determination of the means of processing and at the time of the processing itself. The General Data Protection Regulation defines a list of principles related to personal data processing (cf. Art. 5) that in particular have to be considered (cf. Chapter 3).

Several researchers and data protection commissioners have published guidelines on how privacy by design could be understood [55, 183, 171, 111, 75, 197]. However, many system developers are not familiar with privacy principles or technologies that implement them. Their work usually focuses on realising functional requirements, where other demands—e.g. privacy or security guarantees—fall short as a result. Also, the developing tools provided by software companies hardly considers privacy principles. Similarly, data protection authorities currently have limited means to evaluate how well an ICT system implements the privacy by design principle. It is not so easy to overcome those shortcomings because there are conceptual difficulties in guaranteeing privacy properties in dynamic systems, i.e., systems that adopt to changing requiremetnts. Still, the degree of implementation of privacy principles in today's ICT landscape and the consideration of privacy requirements in the design process could—and should—be considerably increased—no matter whether this will be demanded by law or not.

### Objectives of this report

On this basis, this report aims to contribute to the discussion on how engineering privacy by design can be made concretely and effectively. We aim to identify the challenges associated with engineering privacy and describe ways to build in privacy in the lifecycle of system development (cf. Chapter 2). This approach is substantiated by discussing privacy design strategies which encompass design patterns, data- and process-oriented design strategies, and privacy-enhancing technologies (cf. Chapter 3). Thereafter we give a structured overview of important privacy technologies (cf. Chapter 4). Further, we outline the policy context related to privacy by design (cf. Annex A: ). Finally, we exemplify limits to privacy by design and issue recommendations that address system developers, service providers, data protection authorities, and policy makers (cf. Chapter5).

This report provides a basis for better understanding of the current state of the art concerning privacy by design. It can be used as repository for system developers who strive to integrate privacy principles in their systems. Further, it can be used as a reference for data protection authorities for their tasks of monitoring the application of the data protection law as well as providing guidance. Lastly, the report should help regulators to better understand the opportunities, challenges and limits of the by-design principles, to improve the expressiveness and effectiveness of future policy.

---

[7] Dr. Ann Cavoukian, former Information and Privacy Commissioner of Ontario, Canada; for PbD see http://www.privacybydesign.ca/

## 2   Engineering Privacy

Privacy by design is a multifaceted concept: in legal documents on one hand, it is generally described in very broad terms as a general principle; by computer scientists and engineers on the other hand it is often equated with the use of specific privacy-enhancing technologies (PETs). However, privacy by design is neither a collection of mere general principles nor can it be reduced to the implementation of PETs. In fact, it is a process involving various technological and organisational components, which implement privacy and data protection principles. These principles and requirements are often derived from law, even though they are often underspecified in the legal sources.

As any process, privacy by design should have well-defined objectives, methodologies, and evaluation means. Currently, the definition of methodologies is not within the scope of the European legal data protection framework, but the upcoming European General Data Protection Regulation provides useful indications with regard to objectives and evaluations of a privacy-by-design process, including data protection impact assessment, accountability and privacy seals.

This chapter, we start by summarising important approaches to privacy engineering and related concepts (Section 2.1). On this basis we present privacy and data protection principles derived from the legal data protection framework with a focus on the European Union (Section 2.2). Then, we provide a general overview of a generalised privacy-by-design process; we define the context and objectives (Section 2.3), methodologies (Section 2.4) and evaluation means (Section 2.5).

### 2.1   Prior art on privacy engineering

The understanding of principles concerning privacy and data protection has evolved over the years—on the international, European and national level[8]. This report deals with informational privacy of individuals, i.e. natural persons, and their demands of protection for their personal data starting from the European legal data protection framework. The "privacy by design" procedures and tools that are depicted and discussed in this report may address only parts of the comprehensive privacy and data protection concept, or they may go beyond, e.g. by protecting personal data not only of individuals, but also of groups, or by supporting anti-censorship requirements.

#### Protection of which data?

The concepts of privacy and data protection must not be reduced to protection of data. In fact, the concepts have to be understood more broadly: they address the protection of human beings and their personal rights as well as democratic values of society. Keeping this in mind, privacy and data protection require safeguards concerning specific types of data since data processing may severely threaten informational privacy.

Several terms have been introduced to describe types of data that need to be protected. A term very prominently used by industry is "personally identifiable information (PII)", i.e., data that can be related to an individual. Similarly, the European data protection framework centres on "personal data". However, some authors argue that this falls short since also data that is not related to a single individual might still have an impact on the privacy of groups, e.g., an entire group might be discriminated with the help of certain information. For data of this category the term "privacy-relevant data" has been used. Evangelists of the latter school express that privacy protection measures should always be applied if risks for privacy and data protection cannot be excluded. In several cases data were—at least in the beginning—not considered as personal data, but later risks for the privacy of individuals or for group discrimination based on such data became apparent. "Privacy-relevant data" also comprises types of data that contribute to or enable linkage of other data sets and thereby establish a relation to a person that has not been visible before, or data that helps to infer sensitive information without linkage. For instance, a collection of timestamps may not be seen as PII or personal data, but they could contribute to data linkage and thereby may have some influence on privacy risks. Note, "privacy-relevant data" is a super group of PII/personal data.

---

[8] Various definitions for privacy, data protection and related concepts have been proposed. This affects the understanding of requirements to be realised and supported in legal, organisational and technical systems. The terminology even diverges in different communities dealing with "privacy by design" or "data protection by design". However, we will not join the to some extent sophisticated, yet justified debate on differences of the various concepts, but focus on similarities to identify overarching principles and approaches.

### Early approaches

Early approaches for a globally or supra-nationally harmonised interpretation of what privacy and data protection should constitute comprise the OECD Guidelines on the Protection of Privacy and Trans-border Flows of Personal Data from 1980 ([151], revised in 2013 [152]) or the Fair Information Practice Principles (FIPPs, [193], based on a report from 1973 [192]) from the United States Federal State Commission. Also, the European Data Protection Directive 95/46/EC [82] from 1995 is based on similar principles (see more on the European policy context in Chapter 6).

Since the 1970s and driven by a rather technical community, principles for guidance of data processing have been proposed.

### Multilateral Security

The concept of Multilateral Security has been discussed since 1994 [162, 163]: Whereas system design very often does not or barely consider the end-users' interests, but primarily focuses on owners and operators of the system, multilateral security demands to take into account the privacy and security interests of all parties involved. To realise that, each party should determine the individual interests as well as privacy and security goals and express them. This requires that all parties be informed about associated advantages (e.g. security gains) and disadvantages (e.g. costs, use of resources, less personalisation). All these interests and goals would be taken into account for the choice of mechanisms to support and realise the demands of all parties.

Multilateral security aims at specifically empowering end-users to exercise their rights and play an active role in deciding on the means of data processing. A prerequisite would be transparent and trustworthy ICT systems. In consequence, the idea of multilateral security could implement personal control and freedom of choice [204] as well as the individual's right to informational self-determination [45] which has become the basis of data protection in Germany after a ruling of the Federal Constitutional Court in 1983 and has influenced the discussion on the European level since.

The approach to multilateral security has been developed in parallel to the first concepts of privacy-enhancing technologies and later the notion of "privacy by design". The research on multilateral security took already into account and was inspired by the work of David Chaum and other researchers on anonymity and data minimisation (e.g. [57, 59, 60]). Similar challenges as with privacy by design have been identified by the multilateral security research community, e.g. the necessity to embed the options for privacy in system design because for ICT systems that involve more than one party multilateral security cannot be achieved unilaterally on the end-user's side. The characteristics and maturity of technical components for multilateral security have been analysed in [156].

The multilateral security approach addresses the advantages of implementing in privacy features at an early stage, at best already in the design phase of conceptualisation. Since the negotiation between the parties involved is key for the concept, security and privacy features to choose from should be built in, but "privacy by default"—at least in the original concept—plays a less important role.

### Privacy-Enhancing Technologies

New information technologies change the privacy and data protection risks we are facing, that is new risks (e.g. through ease of search, cheap data storage) emerge, but technology can also help to minimise or avoid risks to privacy and data protection. Starting in the 1970s the research community and most prominently David Chaum explored the field of privacy technologies. For instance technologies were proposed for anonymous electronic communication [57], transactions ([59], and payment [60]. In 1995, the idea of shaping technology according to privacy principles was discussed among Privacy and Data Protection Commissioners. At the time the main principles were data minimisation and identity protection by anonymisation or pseudonymisation [119, 39]. This discussion led to the term "Privacy-Enhancing Technologies (PETs)" [119, 40].

Development and integration of PETs means built-in privacy and the consideration of the full lifecycle of a system. "Privacy by default" is also addressed, in particular by stressing the data minimisation principle.

### Global Privacy Standards

With regard to the globalisation of business practices, international standards on the protection of personal data and privacy were needed. Thus in 2006 the International Data Protection and Privacy Commissioners Conference agreed on a set of universal privacy principles that should function as a global standard for business practices, regardless of borders [54]. This so-called "Global Privacy Standard" has been aiming at assisting public policy makers as well as businesses and developers of technology. The Global Privacy Standard does not substitute the legal privacy and data protection obligations in the respective jurisdiction where data processing takes place, but is noteworthy because it lists "accountability" as a principle apart and incorporates "data minimisation" as part of the "collection limitation" principle. While none of these was mentioned in the European Data Protection Directive from 1995 and unanimously regarded as important for "privacy by design" in the research community.

A further evolved set of principles was adopted at the International Data Protection and Privacy Commissioners Conference in 2009: the International Standards on the Protection of Personal Data and Privacy—the so-called "Madrid Resolution" [1]—approved by Privacy and Data Protection Authorities of over 50 countries. The standards document defines a series of 20 principles and rights that can be taken as a base for the development of an internationally binding tool to protect individual rights and freedoms at a global level. Representative from large multinational companies have announced to support these principles to promote a worldwide framework for privacy protection. However, in contrast to the Global Privacy Standard, the Madrid Resolution primarily addresses lawmakers by discussing how to install a control regime on processing personal data.

### Privacy by Design

Some may have misunderstood PETs as the panacea that could solve all privacy problems simply by adding PET components on top of an existing system. Of course this is hardly ever the case. Clarification was provided by Ann Cavoukian's introduction of the privacy-by-design approach with its seven foundational principles [55], particularly by demanding privacy to be embedded into design as a preventive and proactive measure.

One year after the Madrid Resolution on worldwide privacy standards, the International Conference of Data Protection and Privacy Commissioners adopted privacy by design "as a holistic concept that may be applied to operations throughout an organisation, end-to-end, including its information technology, business practices, processes, physical design and networked infrastructure" (the so-called "Privacy by Design Resolution" [2]). This resolution states that "existing regulation and policy alone are not fully sufficient to safeguard privacy" and underlines that "embedding privacy as the default into the design, operation and management of ICT and systems, across the entire information life cycle, is necessary to fully protect privacy".

The seven foundational principles are characterising properties rather than instructions for specific measures to be taken. Further explanation of how to operationalise privacy by design is provided in [56]. Related approaches that emphasise the need of practicability have been proposed in [111], integrating requirements analysis and modelling of attackers, threats and risks, or in [183] which distinguishes between "privacy-by-policy" (with focus on notice and choice principle) and "privacy-by-architecture" (with focus on data minimisation).

### The Privacy Principles of ISO/IEC 29100

In 2011, the "Privacy framework" (ISO/IEC 29100, [125]) from the International Organisation for Standardization (ISO) and the International Electrotechnical Commission (IEC) was published as an international standard. It targets organisations and intends to support them in defining their privacy safeguarding requirements. Where legal privacy and data protection requirements exist, the standard is to be seen as complementary. It aims at enhancing today's security standards by the privacy perspective whenever personally identifiable information is processed. The target audience of the framework are mainly organisations as being responsible entities for data processing; in addition ICT system designers and developers are addressed.

Although it was developed within the subcommittee "IT Security techniques" (SC 27), the standard goes beyond IT security by explicitly linking to ISO/IEC 27000 concepts and demonstrating their correspondences to each other. Beside eleven elaborated privacy principles the privacy framework comprises a brief description of a common privacy terminology, the actors and their roles in processing personally identifiable information, privacy safeguarding requirements and controls.

### Privacy Protection Goals

Protection goals aim to provide abstract, i.e. context independent properties for IT systems. In ICT security the triad of *confidentiality*, *integrity,* and *availability* has been widely accepted. While several extensions and refinements have been proposed, these core protection goals remained stable over decades and served as basis for many ICT security methodologies; developers are familiar with them and employ them to structure the identification of risks and choosing appropriate safeguards.

As complement to these security protection goals, three privacy-specific protection goals were proposed in 2009 [172], namely *unlinkability, transparency,* and *intervenability*. These were further refined in [116, 158], and embedded in a standardised data protection model that is being acknowledged by the Data Protection Authorities in Germany and proposed for use on the European level [170, 36].

- *Unlinkability* ensures that privacy-relevant data cannot be linked across domains that are constituted by a common purpose and context, and that means that processes have to be operated in such a way that the privacy-relevant data are unlinkable to any other set of privacy-relevant data outside of the domain.[9]

---

[9] Note that this definition of "unlinkability" focuses on separating domains characterised by purpose and context, as this is being done in many data protection laws. Of course stricter unlinkability realisations (see e.g. [150]) may support this goal as well.

Unlinkability is related to the principles of necessity and data minimisation as well as purpose binding. Mechanisms to achieve or support unlinkability comprise of data avoidance, separation of contexts (physical separation, encryption, usage of different identifiers, access control), anonymisation and pseudonymisation, and early erasure or data.

- *Transparency.* Transparency ensures that all privacy-relevant data processing including the legal, technical and organisational setting can be understood and reconstructed at any time. The information has to be available before, during and after the processing takes place. Thus, transparency has to cover not only the actual processing, but also the planned processing (ex-ante transparency) and the time after the processing has taken place to know what exactly happened (ex-post transparency). The level of how much information to provide and in which way it should be communicated best has to be tailored according to the capabilities of the target audience, e.g. the data controller, the user, an internal auditor or the supervisory authority.

  Transparency is related to the principles concerning openness and it is a prerequisite for accountability. Mechanisms for achieving or supporting transparency comprise logging and reporting, an understandable documentation covering technology, organisation, responsibilities, the source code, privacy policies, notifications, information of and communication with the persons whose data are being processed.

- *Intervenability.* Intervenability ensures intervention is possible concerning all ongoing or planned privacy-relevant data processing, in particular by those persons whose data are processed. The objective of intervenability is the application of corrective measures and counterbalances where necessary.

  Intervenability is related to the principles concerning individuals' rights, e.g. the rights to rectification and erasure of data, the right to withdraw consent or the right to lodge a claim or to raise a dispute to achieve remedy. Moreover, intervenability is important for other stakeholders, e.g. for data controllers to effectively control the data processor and the used IT systems to influence or stop the data processing at any time. Mechanisms for achieving or supporting intervenability comprise established processes for influencing or stopping the data processing fully or partially, manually overturning an automated decision, data portability precautions to prevent lock-in at a data processor, breaking glass policies, single points of contact for individuals' intervention requests, switches for users to change a setting (e.g. changing to a non-personalised, empty-profile configuration), or deactivating an auto pilot or a monitoring system for some time. Note these mechanisms often need the cooperation of the service provider (often referred as honest-but-curious-attacker model).

Working with protection goals means to balance the requirements derived from the six protection goals (ICT security and privacy) concerning data, technical and organisational processes. Considerations on lawfulness, fairness and accountability provide guidance for balancing the requirements and deciding on design choices and appropriate safeguards.

## 2.2   Deriving privacy and data protection principles from the legal framework

The above described principles and mechanisms do need to be reflected in legislation to be effective. In this report we focus on the EU perspective. The main principles from the European legal data protection context are summarised and briefly discussed. Thereby, references to the European Data Protection Directive 95/46/EC (in short: DPD) [82], to Opinions of the Article 29 Data Protection Working Party (based on Art. 29 DPD) and to the proposed European General Data Protection Regulation (in short: GDPR) [159] are given.

## Lawfulness

According to European data protection law, the processing of personal data is only allowed if (a) the individual whose personal data are being processed (in the European legal framework called "data subject") has unambiguously given consent, or processing is necessary (b) for the performance of a contract, (c) for compliance with a legal obligation, (d) in order to protect vital interests of the data subject, (e) for the performance of a task carried out in the public interest, or (f) for the purposes of legitimate interests pursued by the data processing entities if such interests are not overridden by the fundamental rights and freedoms of the data subject.

"Personal data" means any information relating to an identified or identifiable natural person—for a detailed discussion see [19]. This is related to the term personally identifiable information (PII), as e.g. used in the privacy framework standardised by ISO/IEC [125].

This very basic principle of lawfulness is not internationally harmonised, i.e., while in several countries outside Europe processing of personal data is permitted unless it is explicitly forbidden, in the EU processing is usually forbidden unless there is an explicit permission, e.g. by the individual's consent or by statutory provisions.

Note that for legally compliant data processing regulatory norms other than such concerning privacy need to be considered. Some of them contain requirements colliding with well-known privacy and data protection principles, e.g. legally demanded data retention overruling data minimisation considerations.

**References to European data protection law**

- Art. 7 DPD,
- Art. 29 Data Protection Working Party: "Opinion on the concept of personal data" [19],
- Art. 29 Data Protection Working Party: "Opinion on the notion of legitimate interests of the data controller" [22],
- Art. 5(1) point (a) GDPR principles "lawfulness, fairness and transparency", and
- Art. 6 GDPR "lawfulness of processing".

## Consent

The term consent is further specified in the legal framework. To enable lawful data processing of individuals' personal identifiable information, individuals need to give specific, informed and explicit indication of their intentions with respect to the processing of their data. A declaration of consent is invalid if not all these requirements are met. Hence, transparency is a prerequisite for consent. Furthermore, consent can be withdrawn with effect for the future. Consent is related to the right to informational self-determination [45] and by this an expression of the individuals' freedoms in general. However, in practice many individuals are not sufficiently informed, or the consent is not freely given. In the opinion of the authors, this deplorable situation occurs due to two issues, namely the way consent is asked is too complex, and the individuals' focus is on another topic, at the moment consent is asked. This observation has been made not only in the field of privacy and data protection with "take it or leave it" apps or contracts in legalese, but also when signing consent forms for medical measures or bank statements. This has discredited the concept of consent.

**References to European data protection law**

- Art. 2 point (h) DPD,
- Art. 29 Data Protection Working Party "Opinion on the definition of consent" [19],
- Art. 4(8) GDPR definition of the "data subject's consent", and
- Art. 7 GDPR "conditions for consent".

## Purpose binding

Personal data obtained for one purpose must not be processed for other purposes that are not compatible with the original purpose. The purpose has to be legitimate, and it has to be specified and made explicit before collecting personal data.

In many countries outside Europe, the principle of purpose limitation or purpose binding is unknown. Instead, it is encouraged to use data for multiple purposes. Big Data is one of the big trends that incorporates multi-purpose linkage and analysis of data instead of leaving them in separated domains.

### References to European data protection law

- Art. 6(1) points (b)-(e) DPD,
- Art. 29 Data Protection Working Party: "Opinion on purpose limitation" [18],
- Art. 29 Data Protection Working Party: "Opinion on personal data" [19],
- Art. 5(1) point (b) GDPR principle "purpose limitation", also
- Art. 5(1) points (c)-(e) GDPR, and
- Art. 21(2a) GDPR.

## Necessity and data minimisation

Only personal data necessary for the respective purpose may be processed, i.e. in the collection stage and in the following processing stage, personal data has to be fully avoided or minimised as much as possible. Consequently, personal data must be erased or effectively anonymised as soon as it is not anymore needed for the given purpose. Although data minimisation at the earliest stage of processing is a core concept of privacy-enhancing technologies cf. [119, 40], and it has been mentioned explicitly in the Global Privacy Standard of 2006, it has not been well enforced, yet.

### References to European data protection law

- Art. 7 DPD,
- Art. 29 Data Protection Working Party: "Opinion on anonymisation techniques" [16],
- Art. 29 Data Protection Working Party: "Opinion on the application of necessity and proportionality concepts and data protection within the law enforcement sector" [21],
- Art. 5(1) point (c) GDPR principle "data minimisation",
- Art. 5(1) point (e) GDPR principle "storage minimisation", and
- Art. 23 GDPR "data protection by design and by default".

## Transparency and openness

Transparency and openness mean that the relevant stakeholders get sufficient information about the collection and use of their personal data. Furthermore, it needs to be ensured that they understand possible risks induced by the processing and actions they can take to control the processing.

Transparency is a necessary requirement for fair data processing, since (1) individuals need information to exercise their rights, (2) data controllers need to evaluate their processors, and (3) Data Protection Authorities need to monitor according to their responsibilities. Currently, the transparency level is entirely inadequate whereas the complexity of data processing and system interaction is further increasing. However, full transparency might not be possible (nor desirable) due to law enforcement requirements and business secrets.

### References to European data protection law

- Art. 10 DPD, Art. 11 DPD and Art. 14 DPD (obligations to inform the data subject),
- Art. 12 (a) ("right of access"),

- Art. 29 Data Protection Working Party: "Opinion on more harmonised information provisions" [20],
- Art. 5(1) point (a) GDPR (principles "lawfulness, fairness and transparency"),
- Art. 10a GDPR ("general principles for data subject rights"),
- Art. 11 GDPR ("concise, transparent, clear and easily accessible policies"),
- Art. 13a GDPR ("standardised information policies"),
- Art. 14 GDPR ("information to the data subject"),
- Art. 15 ("right to access and to obtain data for the data subject"), and
- Art. 12 (for defining the conditions for exercising data subject rights).

## Rights of the individual

Individuals have right to access and rectify as well as (constrained) to block and erase their personal data. Further they have the right to withdraw given consent with effect for the future. These rights should be supported in a way that individuals can effectively and conveniently exercise their rights.

The implementation, or at least support, of these rights is promoted by the privacy by design principle that demands considering the user and the one that stipulates privacy by default.

### References to European data protection law

- Art. 12 point (b) and (c) DPD ("right of access", in point (b) in particular: the rectification, erasure or blocking of data if appropriate),
- Art. 14 DPD ("right to object"),
- Art. 5 No. 1 (ea) GDPR (principle "effectiveness"),
- Art. 7(3) GDPR (right to withdraw consent at any time),
- Art. 10a GDPR ("general principles for data subject rights"),
- Art. 13 GDPR ("notification requirement in the event of rectification and erasure"),
- Art. 17 GDPR ("right to erasure"),
- Art. 19 GDPR ("right to object"), and
- Art. 12 (for defining the conditions for exercising data subject rights).

## Information security

Information security addresses the protection goals confidentiality, integrity, availability. All of these goals are important also from a privacy and data protection perspective that specifically requires that unauthorised access and processing, manipulation, loss, destruction and damage are prevented. Further, the data have to be accurate. Moreover, the organisational and technical processes for appropriately handling the data and providing the possibility for individuals to exercise their rights have to be available whenever necessary. This principle calls for appropriate technical and organisational safeguards.

### References to European data protection law

- Art. 16 DPD "Confidentiality of processing",
- Art. 17 DPD "Security of processing",
- Art. 5 No. 1 (d) GDPR (principle "accuracy"),
- Art. 5 No. 1 (ea) GDPR (principle "integrity"),
- Art. 30 GDPR "Security of processing", and
- Art. 50 GDPR ("Professional secrecy").

**Accountability**

Accountability means to ensure and to be able to demonstrate the compliance with privacy and data protection principles (or legal requirements). This requires clear responsibilities, internal and external auditing and controlling of all data processing. In some organisations, Data Protection Officers are installed to perform internal audits and handle complaints. A means for demonstrating compliance can be a data protection impact assessment.

On a national level, accountability is supported by independent Data Protection Authorities for monitoring and checking as supervisory bodies.

**References to European data protection law**

- In the DPD, accountability is not directly stated, but aspects of the principle can be seen, among others, in Art. 17 DPD (Security of processing) or by mentioning the possibility of appointing a "personal data protection official" in Art. 18 DPD who should be responsible for ensuring the application of data protection law.
- Art. 29 Data Protection Working Party: "The Future of Privacy" [23],
- Art. 29 Data Protection Working Party: "Opinion on the principle of accountability" [17],
- Art. 5(1) point (f) GDPR and Art. 22 GDPR ("Responsibility and accountability of the controller"),
- Art. 33 GDPR ("Data protection impact assessment"), and
- Art. 35 GDPR ("Designation of the data protection officer").

**Data protection by design and by default**

The principle "Privacy/data protection by design" is based on the insight that building in privacy features from the beginning of the design process is preferable over the attempt to adapt a product or service at a later stage. The involvement in the design process supports the consideration of the full lifecycle of the data and its usage.

The principle "Privacy/data protection by default" means that in the default setting the user is already protected against privacy risks. This affects the choice of the designer which parts are wired-in and which are configurable [117]. In many cases, a privacy-respecting default would not allow an extended functionality of the product, unless the user explicitly chooses it.

**References to European data protection law**

- In the DPD, data protection by design is rather indirectly addressed, e.g. in Art. 17 DPD (Security of processing) where appropriate safeguards are demanded, even if this provision was mainly directed to security instead of privacy guarantees.
- Art. 29 Data Protection Working Party: "The Future of Privacy" [23], and
- Art. 23 GDPR ("Data protection by design and by default").

## 2.3 Definition of the context and objectives

As discussed in the introduction, the integration of privacy requirements in the design of a system is not a simple task. First privacy in itself is a complex, multifaceted and contextual notion. In addition, it is generally not the primary requirement of a system and it may even come into conflict with other (functional or non-functional) requirements. It is therefore of paramount importance to define precisely the goals of a privacy by design process. These goals should form the starting point of the process itself and the basis of its evaluation. One way to define the objectives of the system in terms of privacy is to conduct a preliminary Privacy Impact Assessment (PIA) or a privacy risk analysis. In fact,

PIAs or "data protection impact assessment"—we use both expression interchangeably in this document—are required in certain situations by the GDPR which stipulates that its results should be taken into account in the privacy by design process, cf. GDPR, Paragraph 1 of Article 23; "Data protection by design shall have particular regard to the entire lifecycle management of personal data from collection to processing to deletion, systematically focusing on comprehensive procedural safeguards regarding the accuracy, confidentiality, integrity, physical security and deletion of personal data. Where the controller has carried out a data protection impact assessment pursuant to Article 33, the results shall be taken into account when developing those measures and procedures."

Recently considerable efforts have been put on PIAs, namely, several handbooks have been published and more and more PIAs are conducted in different countries. The interested reader can find an outline of the current situation in [206] and more detailed accounts in [206, 207]. A PIA can be a more or less heavy process, including different organisational, decision making and technical tasks, with PIA reports varying from a page and a half to more than one hundred and fifty pages. From a technical point of view, the core steps of a PIA are:

1. the identification of stakeholders and consulting of these stakeholders,
2. the identification of risks (taking into account the perception of the stakeholders),
3. the identification of solutions and formulation of recommendations,
4. the implementation of the recommendations,
5. reviews, audits and accountability measures.

The inputs of the privacy by design process per se should be the outputs of the second step (risk analysis) and third step[10] (recommendations) and its output contributes to step 4 (implementation of the recommendations[11]). In fact, it could be argued that privacy by design actually encompasses the entire process (including the PIA itself), but we focus on the actual design phase in this document. Regardless of the chosen definitions, it should be emphasised that privacy by design is an iterative, continuous process and PIAs can be conducted at different stages of this process.

The key step of a PIA with respect to privacy by design is therefore the risk analysis. A wide range of methods have been defined for IT security risk analysis [137] but much less are dedicated to privacy risk analysis [66, 75, 150]. The OASIS Privacy Management Reference Model and Methodology (PMRM) [44] includes a general methodology for analysing privacy policies and their privacy management requirements but it does not provide a precise risk analysis method. The only few existing privacy risk analysis methods are actually adaptations or transpositions of security risk analyses to privacy, for example the EBIOS method in [66] or STRIDE in [75].

As an illustration, the core of the risk analysis method published by the French Commission Nationale Informatique et Libertés (CNIL) is the identification of the "feared events" (the events that should be prevented) and the threats (what can make the feared events happen). Examples of feared events can be the illegal use of personal information for advertising emails or the failure of the data controller to comply with the deletion obligation leading to outdated or incorrect data that can be used to deprive subjects from their rights (employment, social benefits, etc.). The level of severity of a feared event is derived from two parameters: the level of identification of the personal data (how easy it is to identify data subjects from the asset at risk) and the level of damage (potential impact of the feared event for the subject). As far as threats are concerned, their likelihood is calculated from the vulnerabilities of the supporting assets (to what extent the features of the supporting assets can be exploited in order to carry out a threat) and the capabilities of the risk sources (attackers) to exploit these vulnerabilities (skills, available time, financial resources, closeness to the system, motivation, etc.). Each privacy risk consisting of a feared event and the associated threats can then be plotted in a two-dimensional (likelihood and severity) space. Depending on its position in this space, a risk can be classified as "to be absolutely avoided", "to be mitigated" (to reduce its likelihood and/or severity), or "acceptable" (very unlikely and with minor impact). The CNIL method itself is rather general and high-level but it is supplemented by a catalogue of good practices to help data controllers in their task (to evaluate the impact of the threat events, to identify the sources of risks, to select measures proportionate to the risks, etc.).

---

[10] Step 3 can be a set of general recommendations about the system rather than a precise, well-defined design. A PIA can also be applied to an already existing system or a system whose architecture has already been defined, in which case privacy by design and risk analysis are more interwoven.
[11] Some of the recommendations of a PIA are not technical.

The LINDDUN methodology[12] broadly shares the principles of the CNIL method but it puts forwards a more systematic approach based on data flow diagrams and privacy threat tree patterns. The first step of the method consists in building a data flow diagram providing a high-level description of the system. A data flow diagram involves entities (e.g. users), processes (e.g. a social network service), data stores (e.g. database) and data flows between the entities. The second step consists in mapping privacy threats to components of the data flow diagram (e.g. associating the "identifiability of the subject" threat to the "database" component). The methodology considers seven types of threats[13]: linkability, identifiability of the subject, non-repudiation, detectability of an item of interest, information disclosure, content unawareness, policy, and consent non-compliance. The third step is the identification of more precise misuse scenarios. To this aim, LINDDUN provides a catalogue of threat tree patterns showing the preconditions for a privacy attack scenario (in the form of AND/OR trees). A precondition is typically a vulnerability that can be exploited by an attacker having the necessary means to conduct an attack. Privacy requirements can then be elicited from the misuse cases. Just like in the CNIL methodology, the severity level of the privacy threats has to be evaluated, risks have to be prioritised and privacy requirements can be addressed in different ways (ranging from removing a feature of the system, warning the user, taking organisational measures and using privacy-enhancing technologies). The choice of appropriate privacy-enhancing technologies is facilitated by correspondence tables associating privacy objectives and data flow components with PETs.

Whatever method is used, a risk analysis should therefore provide useful inputs to the design process: assumptions on the context (including the external environment, the attackers and their capabilities), precise privacy objectives for the system, and potentially suggestions of privacy-enhancing technologies. Even if no formal PIA or risk analysis methodology is conducted, these inputs should be defined at the very beginning of the process since otherwise the objectives of the privacy by design process itself would not be well-defined.

## 2.4  Methodologies

When the objectives and assumptions on the context are defined, the challenge for designers is to find the appropriate privacy-enhancing techniques and protocols and combine them to meet the requirements of the system. A first hurdle to be overcome are the potential conflicts or inconsistencies between privacy objectives and the other (functional and non-functional) requirements of the system. Another source of complexity is the fact that there exists a huge spectrum of PETs, their specifications can be subtle and understanding how to combine them to achieve a given purpose is cumbersome and error prone. We discuss PETs in Chapter 5 in more detail.

For the above reasons, the development of appropriate methodologies or development tools for privacy by design has been advocated or studied by several research groups during the last decade [14, 13, 111, 120, 131, 44, 187]. A way to face the complexity of the task is to define the privacy by design methodology at the level of architectures [14, 135]. Indeed, most of the reasons identified in [27] why architectures matter are relevant for privacy by design [14]: first, they are the carriers of the earliest and hence most fundamental hardest-to-change design decisions; in addition, they reduce design and system complexity because they make it possible to abstract away unnecessary details and to focus on critical issues. Architectures can therefore help designers reason about privacy requirements and the combination of PETs which could be used to meet them.

---

[12] https://distrinet.cs.kuleuven.be/software/linddun/
[13] Which is the origin of the mnemonic LINDDUN acronym.

If the choice is made to work at the architectural level, the next question is: "how to define, represent and use architectures?" In practice, architectures are often described in a pictorial way, using different kinds of graphs with legends defining the meaning of nodes and vertices, or semi-formal representations such as UML diagrams (class diagrams, use case diagrams, sequence diagrams, communication diagrams, etc.). When stronger guarantees are needed, it is also possible to rely on formal (mathematical) methods to prove that, based on appropriate assumptions on the PETs involved, a given architecture meets the privacy requirements of the system.

Whatever the chosen level of description and representation language for the system, the following key criteria have to be considered in a privacy by design methodology:

- *Trust assumptions*. A key decision which has to be made during the design phase is the choice of the trust relationships between the stakeholders: this choice is a driving factor in the selection of architectural options and PETs. Any disclosure of personal data is conditional upon a form of trust between the discloser and the recipient. Different types of trust can be distinguished such as blind trust, verifiable trust, and verified trust [14, 13]. Blind trust is the strongest form of trust; from a technical point of view it could lead to the weakest solutions, the ones most vulnerable to misplaced trust. Verifiable trust falls between the two other options: trust is granted by default but verifications can be carried out a posteriori (for example using commitments and spot checks) to check that the trusted party has not cheated. In contrast, verified trust amounts, technically speaking, to a "no trust" option; it relies on cryptographic algorithms and protocols (such as zero knowledge proofs, secure multiparty computation or homomorphic encryption) to guarantee, by construction, the desired property. Furthermore the amount of trust necessary can be reduced by the use of cryptographic techniques or distribution of data. Other forms of trust, involving groups of stakeholders, can also be considered. For instance, distributed trust is a form of trust that depends on the assumption that the data is split between several entities that are assumed not to collude (or at least a minimum fraction of them).

- *Involvement of the user.* The second batch of decisions to be taken by the designer is the types of interactions with the users. For some systems (smart metering, electronic traffic pricing, etc.) it may be the case that no interaction with the user is necessary to get his consent (which is supposed to have been delivered through other, non-technical means, or which is not required because it is not the legal ground for the collection of the data) or to allow him to exercise his rights. In other cases, these interactions have to be implemented, which means that several questions have to be addressed by the designer: what information is communicated to the user, in what form and at what time(s)? What initiatives can the user take, through what means, at what time(s)? Great care should be taken that the interface of the system allows subjects to exercise all their rights (informed consent, access, correction, deletion, etc.) without undue constraints.

- *Technical constraints.* Some constraints on the environment usually have to be taken into account such as, for example, the fact that a given input data is located in a specific area, provided by a sensor that may have limited capacities, or the existence (or lack of) communication channel between two components.

- *Architecture.* The answers to the above questions usually make it possible to significantly reduce the design space. The last stage is then the definition of the architecture, including the type of components used, the stakeholders controlling them, the localisation of the computations, the communication links and information flows between the components.

This methodology should be supported by appropriate libraries of privacy design strategies and PETs to help the designers in the last "creative" step (choice of the architecture and components). Existing privacy by design strategies and PETs are presented in Chapter 4 and Chapter 5 respectively.

The privacy by design approach is a continuous, iterative process [44] and various events (such as the availability of new PETs or attacks on existing technologies) can call for a reordering of priorities or reconsideration of certain assumptions. As for any process, it is also necessary to be able to evaluate its result as discussed in the next section.

## 2.5 Evaluation means

The application of a well-defined privacy by design process is not by itself an absolute guarantee that the system will comply with all its privacy requirements. In addition, the accountability principle stipulates that data controllers must be able to demonstrate compliance to internal and external auditors. This obligation is enshrined in the GDPR: "Impact assessments can only be of help if controllers make sure that they comply with the promises originally laid down in them. Data controllers should therefore conduct periodic data protection compliance reviews demonstrating that the data processing mechanisms in place comply with assurances made in the data protection impact assessment. It should further demonstrate the ability of the data controller to comply with the autonomous choices of data subjects. In addition, in case the review finds compliance inconsistencies, it should highlight these and present recommendations on how to achieve full compliance."[14] This demonstration can take place at different times and in different ways. To clarify this, Colin Bennett has distinguished three layers of accountability [33]: accountability of policy, accountability of procedures, and accountability of practice. The first two types of accountability amount to have appropriate documents available defining the privacy policies implemented by the system, i.e. documentation of privacy requirements[15] and the internal mechanisms and procedures in place (such as PIA, complaint handling procedure, staff training, existence of a privacy officer, etc.) need to be in place. The third type of accountability is more demanding: to comply with accountability of practice, data controllers must be able to demonstrate that their actual data handling complies with their obligations. To do so, it is generally necessary to keep audit logs and these logs must also comply with other privacy principles such as the data minimisation principle (only the personal data necessary for the audits should be recorded) and security obligation (audit logs should not represent an additional source of risks for personal data). Solutions have been proposed to define the information to be recorded in the logs [48, 47] and to ensure their security [178, 30, 203].

Privacy certification or "privacy seals", which are also promoted by the GDPR[16], provide another framework for privacy assessment. Privacy seals have been characterised as follows in [169]: "A pri-vacy seal is a certification mark or a guarantee issued by a certification entity verifying an organisation's adherence to certain specified privacy standards. A privacy seal is a visible, public indication of an organisation's subscription to established, largely voluntary privacy standards that aim to promote consumer trust and confidence in e-commerce". In fact, privacy seals and certificates are already in operation, with different targets: some of them apply to websites[17], others to procedures[18], other to products, etc. Regardless of their target, the most important criteria to establish the value of a privacy seal are the following:

- What is the certification entity and what are the mechanisms in place to ensure that it is trustworthy? The ideal situation is when an independent, official body issues the certificate (even though the evaluation itself can be conducted by accredited laboratories such as in the Common Criteria for security[19]).
- What is the scope of the certification? In the context of privacy by design, we are mostly interested in the certification of products but even then, the precise scope of the evaluation has to be defined (a product can be part of a system) as well as the assumptions on its environment (stakeholders, expected use, etc.).
- What are the privacy requirements or standards used for the evaluation? Each privacy seal subscribes to its own standards or criteria (for example the EuroPriSe criteria[20]) which can be more or less demanding. If a PIA or a risk analysis has been conducted, it should provide (or be included into) the privacy requirements of the product, and a privacy seal that would not take such requirements into account would not provide sufficient guarantees. In any case, the reference for the evaluation should be precisely documented.
- How is the result of the certification communicated to the users? This issue is especially critical when the seal or certificate is supposed to provide information to end-users such as in the case of websites [189]. For more business oriented certificates, the result of the evaluation should describe without ambiguity the scope of the evaluation, the privacy requirements, the level of assurance and the result of the evaluation (which may include observations or recommendations about the use of the product).

---

[14] Paragraph 74(a) of the Recitals

[15] These requirements should be as defined by the PIA or risk analysis process

[16] "In order to enhance transparency and compliance with this Regulation, the establishment of certification mechanisms, data protection seals and standardised marks should be encouraged, allowing data subjects to quickly, reliably and verifiably assess the level of data protection of relevant products and services. A "European Data Protection Seal" should be established on the European level to create trust among data subjects, legal certainty for controllers, and at the same time export European data protection standards by allowing non-European companies to more easily enter European markets by being certified." (Article 77 of the Recitals)

In practice, the added value of a privacy seal or certificate should be an increased confidence in the fact that a product meets its privacy requirements. This increased confidence ultimately depends on the trust that can be placed in the certification body and history has shown that this trust is not always well deserved. As shown by several authors, this can have a counterproductive effect in terms of privacy because privacy seals can create an "illusion of privacy" leading users to disclose more information than they would do in the absence of a seal [169]. Different models of co-regulation are suggested in [169] to avoid this deceptive effect and create an appropriate environment for future privacy seals.

# 3 Privacy Design Strategies

## 3.1 Software design patterns, strategies, and technologies[21]

In this section we will develop the notion of a design strategy, and explain how it differs from both a design pattern and a (privacy-enhancing) technology.

---

[17] The interested reader can find in [180] an overview of online security and privacy seals.
[18] http://www.cnil.fr/linstitution/labels-cnil/procedures-daudit/
[19] https://www.commoncriteriaportal.org/
[20] https://www.european-privacy-seal.eu/EPS-en/Home
[21] This section is based on [118].

Software architecture involves the set of significant decisions about the organisation of a software system[22]. It includes the selection of the structural elements, the interfaces by which a system is composed from these elements, the specified behaviour when those elements collaborate, the composition of these structural and behavioural elements into a larger subsystem, and the architectural style that guides this organisation.

Several software development methodologies exist. In the waterfall model, software system development proceeds in six phases: concept development, analysis, design, implementation, testing and evaluation. Systems are never developed in one go. Typically several iterations are involved, both before and after an initial version is released to the public. Therefore, software development in practice proceeds in a cycle, where after evaluation a new iteration starts by updating the concept as appropriate.

To support privacy by design throughout the software development each of these phases rely on different concepts. In the concept development and analysis phases so called privacy design strategies (defined further on) are necessary. The known concept of a design pattern is useful during the design phase, whereas concrete (privacy-enhancing) technologies can only be applied during the implementation phase.

### 3.1.1 Design patterns

Design patterns are useful for making design decisions about the organisation of a software system. A design pattern "*provides a scheme for refining the subsystems or components of a software system, or the relationships between them. It describes a commonly recurring structure of communicating components that solves a general design problem within a particular context.*" [46]

This allows a system designer to solve a problem by breaking it down into smaller, more manageable, sub-problems, without bogging the designer with implementation details. At the same time, the pattern clearly describes the consequences of the proposed subdivision, allowing the designer to determine whether the application of that pattern achieves the overall goal. This is why the description [104] of a design pattern typically contains the following elements: its name, purpose, context (the situations in which it applies), implementation (its structure, components and their relationships), and the consequences (its results, side effects and trade-offs when applied). An example of a software design pattern is the Model-View-Controller[23] that separates the representation of the data (the model) from the way it is presented towards the user (the view) and how the user can interact with that data (using the controller).

In the context of privacy by design, few design patterns have been explicitly described as such to date. We are aware of the work of Hafiz [113, 114], Pearson [155, 154], van Rest et al. [197], and a recent initiative of the UC Berkeley School of Information[24]. Many more implicit privacy design patterns exist though, although they have never been described as such. We will encounter some of them a bit further in this report.

### 3.1.2 Design strategies

Many design patterns are very specific, and therefore cannot be applied directly in the concept development phase. However, there exist design patterns of a higher level of abstraction, called architecture patterns. Such architecture patterns "*express a fundamental structural organisation or schema*

---

[22] Based on an original definition by Mary Shaw, expanded in 1995 by Grady Booch, Kurt Bittner, Philippe Kruchten and Rich Reitman as reported in [130].

[23] Originally formulated in the late 1970s by Trygve Reenskaug at Xerox PARC, as part of the Smalltalk system.

[24] http://privacypatterns.org/

*for software systems. They provide a set of predefined subsystems, specify their responsibilities, and include rules and guidelines for organising the relationships between them.*"[25]

Some people consider the Model-View-Controller pattern introduced above such an architecture pattern. The distinction between an architecture pattern and a design pattern is not always easily made, however. It does show that one can look at system design at varying levels of abstraction, and that concrete design patterns have been too specific to be readily applicable in certain cases. In fact, there are even more general principles that guide the system architecture. We choose, therefore, to express such higher level abstractions in terms of design strategies. These are defined as follows [120].

A design strategy describes a fundamental approach to achieve a certain design goal. It favours certain structural organisations or schemes over others. It has certain properties that allow it to be distinguished from other (fundamental) approaches that achieve the same goal.

Whether something classifies as a strategy very much depends on the universe of discourse, and in particular on the exact goal the strategy aims to achieve. In general we observe that a privacy design strategy is a design strategy that achieves (some level of) privacy protection as its goal.

Design strategies do not necessarily impose a specific structure on the system although they certainly limit the possible structural realisations of it. Therefore, they are also applicable during the concept development and analysis phase of the development cycle[26].

### 3.1.3    Privacy-enhancing technologies

When talking about privacy by design, Privacy-Enhancing Technologies (PETs) are well known, and have been studied in depth for decades now. (This large body of knowledge is summarised further on in this report.) A well-known definition, and one that was later adopted almost literally by the European Commission [67], was given by Borking and Blarkom et al. [39, 195] as follows:

"*Privacy-Enhancing Technologies is a system of ICT measures protecting informational privacy by eliminating or minimising personal data thereby preventing unnecessary or unwanted processing of personal data, without the loss of the functionality of the information system.*"

In principle, PETs are used to implement a certain privacy design pattern with concrete technology. For example, both 'Idemix' [49] and 'U-Prove' [42] are privacy-enhancing technologies implementing the (implicit) design pattern anonymous credentials. There are many more examples of privacy-enhancing technologies, like 'cut-and-choose' techniques [60], 'onion routing' [57] to name but a few.

## 3.2    Eight privacy design strategies

We will now briefly summarise the eight privacy design strategies as derived by Hoepman [120] from the legal principles underlying data protection legislation. This work distinguishes data oriented strategies (described first) and process oriented strategies (described after that).

### 3.2.1    Data oriented strategies

The following four data oriented strategies can support the unlinkability protection goal (see Section 2.1) and primarily address the principles of necessity and data minimisation (see Section 2.2).

---

[25] See http://best-practice-software-engineering.ifs.tuwien.ac.at/patterns.html, and The Open Group Architecture Framework (TOGAF) http://pubs.opengroup.org/architecture/togaf8-doc/arch/chap28.html

[26] We note that the notion of a privacy design strategy should not be confused with the foundational principles of Cavoukian [57] or the concept of a privacy principle from the ISO 29100 Privacy framework [122].

### Strategy #1: MINIMISE

The most basic privacy design strategy is MINIMISE, which states that the amount of personal data that is processed[27] should be restricted to the minimal amount possible.

This strategy is extensively discussed by Gürses et al. [111]. By ensuring that no, or no unnecessary, data is collected, the possible privacy impact of a system is limited. Applying the MINIMISE strategy means one has to answer whether the processing of personal data is proportional (with respect to the purpose) and whether no other, less invasive, means exist to achieve the same purpose. The decision to collect personal data can be made at design time and at run time, and can take various forms. For example, one can decide not to collect any information about a particular data subject at all. Alternatively, one can decide to collect only a limited set of attributes.

### Design patterns

Common design patterns that implement this strategy are 'select before you collect' [126], 'anonymisation and use pseudonyms' [157].

### Strategy #2: HIDE

The second design strategy, HIDE, states that any personal data, and their interrelationships, should be hidden from plain view. The rationale behind this strategy is that by hiding personal data from plain view, it cannot easily be abused. The strategy does not directly say from whom the data should be hidden. And this depends on the specific context in which this strategy is applied. In certain cases, where the strategy is used to hide information that spontaneously emerges from the use of a system (communication patterns for example), the intent is to hide the information from anybody. In other cases, where information is collected, stored or processed legitimately by one party, the intent is to hide the information from any other party. In this case, the strategy corresponds to ensuring confidentiality.

The HIDE strategy is important, and often overlooked. In the past, many systems have been designed using innocuous identifiers that later turned out to be privacy nightmares. Examples are identifiers on RFID tags, wireless network identifiers, and even IP addresses. The HIDE strategy forces one to rethink the use of such identifiers. In essence, the HIDE strategy aims to achieve unlinkability and unobservability [157]. Unlinkability in this context ensures that two events cannot be related to one another (where events can be understood to include data subjects doing something, as well as data items that occur as the result of an event).

### Design patterns

The design patterns that belong to the HIDE strategy are a mixed bag. One of them is the use of encryption of data (when stored, or when in transit). Other examples are mix networks [57] to hide traffic patterns [57], or techniques to unlink certain related events like attribute based credentials [49], anonymisation and the use of pseudonyms. Techniques for computations on private data implement the HIDE strategy while allowing some processing. Note that the latter two patterns also belong to the MINIMISE strategy.

### Strategy #3: SEPARATE

The third design strategy, SEPARATE, states that personal data should be processed in a distributed fashion, in separate compartments whenever possible.

---

[27] For brevity, and in line with Article 2 of the European directive [83], we use data processing to include the collection, storage and dissemination of that data as well.

By separating the processing or storage of several sources of personal data that belong to the same person, complete profiles of one person cannot be made. Moreover, separation is a good method to achieve purpose limitation. The strategy of separation calls for distributed processing instead of centralised solutions. In particular, data from separate sources should be stored in separate databases, and these databases should not be linked. Data should be processed locally whenever possible, and stored locally if feasible as well. Database tables should be split when possible. Rows in these tables should be hard to link to each other, for example by removing any identifiers, or using table specific pseudonyms.

These days, with an emphasis on centralised web based services this strategy is often disregarded. However, the privacy guarantees offered by peer-to-peer networks are considerable. Decentralised social networks like Diaspora[28] are inherently more privacy-friendly than centralised approaches like Facebook and Google+.

**Design patterns**

No specific design patterns for this strategy are known.

**Strategy #4: AGGREGATE**

The fourth design pattern, AGGREGATE, states that Personal data should be processed at the highest level of aggregation and with the least possible detail in which it is (still) useful.

Aggregation of information over groups of attributes or groups of individuals, restricts the amount of detail in the personal data that remains. This data therefore becomes less sensitive if the information is sufficiently coarse grained, and the size of the group over which it is aggregated is sufficiently large. Here coarse grained data means that the data items are general enough that the information stored is valid for many individuals hence little information can be attributed to a single person, thus protecting its privacy.

**Design patterns**

Examples of design patterns that belong to this strategy are the following: aggregation over time (used in smart metering), dynamic location granularity (used in location based services), *k*-anonymity [185], differential privacy [93] and other anonymization techniques.

### 3.2.2 Process oriented strategies

**Strategy #5: INFORM**

The INFORM strategy corresponds to the important notion of transparency (see also the transparency protection goal in Section 2.1 as well as the principle "transparency and openness" in Section 2.2). Data subjects should be adequately informed whenever personal data is processed.

Whenever data subjects use a system, they should be informed about which information is processed, for what purpose, and by which means. This includes information about the ways the information is protected, and being transparent about the security of the system. Providing access to clear design documentation is also a good practice. Data subjects should also be informed about third parties with which information is shared. And data subjects should be informed about their data access rights and how to exercise them.

---

[28] http://diasporafoundation.org/

**Design patterns**

A possible design patterns in this category is the Platform for Privacy Preferences (P3P)[29]. Data breach notifications are also a design pattern in this category. Finally, Graf et al. [110] provide an interesting collection of privacy design patterns for informing the user from the Human Computer Interfacing perspective. More transparency-enhancing techniques are being mentioned in Section 4.11.

### Strategy #6: CONTROL

The control strategy states that data subjects should be provided agency over the processing of their personal data (see also the privacy protection goal "intervenability" introduced in Section 2.1 as well as the principle of supporting the rights of the individual (Section 2.2).

The CONTROL strategy is in fact an important counterpart to the INFORM strategy. Without reasonable means of controlling the use of one's personal data, there is little use in informing a data subject about the fact that personal data is collected. Of course the converse also holds: without proper information, there is little use in asking consent. Data protection legislation often gives the data subject the right to view, update and even ask the deletion of personal data collected about her. This strategy underlines this fact, and design patterns in this class give users the tools to exert their data protection rights.

CONTROL goes beyond the strict implementation of data protection rights, however. It also governs the means by which users can decide whether to use a certain system, and the way they control what kind of information is processed about them. In the context of social networks, for example, the ease with which the user can update his privacy settings through the user interface determines the level of control to a large extent. So user interaction design is an important factor as well. Moreover, by providing users direct control over their own personal data, they are more likely to correct errors. As a result the quality of personal data that is processed may increase.

**Design patterns**

User centric identity management and end-to-end encryption support control. Furthermore, we refer to Section 4.12 where tools are suggested when discussing intervenability.

### Strategy #7: ENFORCE

The seventh strategy, ENFORCE, states: A privacy policy compatible with legal requirements should be in place and should be enforced. This relates to the accountability principle (see Section 2.2).

The ENFORCE strategy ensures that a privacy policy is in place. This is an important step in ensuring that a system respects privacy during its operation. Of course the actual level of privacy protection depends on the actual policy. At the very least it should be compatible with legal requirements. As a result, purpose limitation is covered by this strategy as well. More importantly though, the policy should be enforced. This implies, at the very least, that proper technical protection mechanisms are in place that prevent violations of the privacy policy. Moreover, appropriate governance structures to enforce that policy must also be established.

**Design patterns**

Access control is an example of a design patterns that implement this strategy. Another example are sticky policies and privacy rights management: a form of digital rights management involving licenses to personal data.

---

[29] http://www.w3.org/P3P/

**Strategy #8: DEMONSTRATE**

The final strategy, DEMONSTRATE, requires a data controller to be able to demonstrate compliance with the privacy policy and any applicable legal requirements. This strategy supports the accountability principles of Section 2.2.

This strategy goes one step further than the ENFORCE strategy in that it requires the data controller to prove that it is in control. This is explicitly required in the new draft EU privacy regulation [159]. In particular this requires the data controller to be able to show how the privacy policy is effectively implemented within the IT system. In case of complaints or problems, she should immediately be able to determine the extent of any possible privacy breaches.

**Design patterns**

Design patterns that implement this strategy are, for example, privacy management systems [50], and the use of logging and auditing.

# 4 Privacy Techniques

## 4.1 Authentication

### 4.1.1 Privacy features of authentication protocols

User authentication is the process by which users in a computer system are securely linked to principals that may access confidential information or execute privileged actions. Once this link is securely established communications can proceed on the basis that parties know each other's identity and a security policy can be implemented. Authentication is key to securing computer systems and is usually the very first step in using a remote service or facility, and performing access control. Strong authentication may also be a key privacy mechanism when used to ensure that only a data subject, or authorised parties, may access private information.

While users may wish to identify themselves, serious privacy concerns could arise from the specific ways in which such authentication is performed. One privacy threat is that a passive network observer, in case of a network authentication protocol, is able to observe users' authentication sessions in order to identify or track them. Another risk, is that the user attempts to authenticate to the wrong service controlled by a malicious entity, and as a result leaks their identity to them (and possibly also their credentials, in so called "phishing attacks" [78]). Finally, successful authentication may lead to a session being established and maintained between two end points. However, a passive observer may again, through intercepting traffic, use the session meta-data to infer the identities of the communicating parties.

State-of-the-art authentication protocols provide protections against attacks outlined above. In particular they prevent third parties from inferring the identities of authenticating parties, do not leak those identities through impersonation, and cannot infer the identity of parties in a secure session.

An example authentication protocol with state-of-the art privacy features is Just Fast Keying (JFK) [7] that builds on ISO 9798-3. The JFK protocol comes in two variants, JFKi providing initiator privacy, and JFKr providing responder privacy. Upon completion both protocols provide communicating parties assurance that they are talking to the claimed entity through the use of certificates and public key cryptography. They also derive a fresh session key to protect the authenticity, integrity and confidentiality of further messages. All variants of JFK have been designed to resist resource depletion attacks and provide forward secrecy (a privacy property we discuss in Section 4.3.3).

The specific privacy features of the protocol are characteristic of the state of the art. Both JFKi/r variants provide privacy with respect to third party passive adversaries. This means that a third party observing a successful or unsuccessful authentication session is unable to infer the identities of either the initiator of the authentication session, or the responder. As a result it is possible for two parties to bootstrap an authenticated relationship without any passive eavesdroppers being able to infer their identities. The two variants of the JFK protocol extend this property to active adversaries: JFKi protects the identity of the initiator of the authentication session, and JFKr protects the identity of the responder. This means that the initiator and responder respectively are not uncovered even if an adversary actively tries to mislead them by initiating or responding with fake messages.

A weaker form of privacy-friendly authentication is common for web-services, and is based on a username and password being disclosed to a service though an encrypted connection [37] (e.g. using the TLS protocol). This mechanism's advantage is its use of standard web technologies, including HTTPS and web forms. It also hides the identity of the client authenticating themselves. However, it does not by default hide the identity of the server, even from a passive observer; a confused user subject to a phishing attack may disclose their identity and credentials (phishing attack); and, in some cases the remaining of the session reverts to unencrypted HTTP, leaking both the identity of the authenticated user and cookies that may be used to impersonate them. As such, this mechanism should be used with great care, and recommend all authenticated interactions be encrypted.

Authentication protocols providing privacy against passive third parties for both initiator and responder, and privacy for one of them in case of an active adversary should be considered the current state-of-the-art. In some circumstances protecting the identity of both initiator and responder against active adversaries is necessary. Protocols achieving this property are known as secret handshake protocols [191], since both parties attempt to hide their identity from the other until they are convinced of parts of the identity of the other—a technically expensive property to achieve. Practical secret handshake protocols have been proposed that allow users to successfully authenticate if both are part of a certain group (sharing a key), that otherwise do not learn any information about the identity of the users or their group membership [53]. Extensions that allow users to privately authenticate in groups have also been proposed.

### 4.1.2 The benefits of end-to-end authentication

The ability to authenticate users is key to providing strong forms of privacy protection for the contents of communications. Public key cryptography may be used to bootstrap a private session key between two parties from public keys that are authenticated as belonging to the two parties. However, the quality and degree of evidence available to users may be variable, depending on the protocols and architectures used. Two key paradigms have emerged:

- Client-server authentication ensures a server, or in general a third party, is assured that the user connected to it is who they claim to be. As a result users are traditionally assigned an identifier (e.g. username) and others using the service can identify them through this identifier. This is, for example, the established architecture for the federated chat protocols XMPP [176], where users authenticate to their XMPP server, which then attests their identity.
- End-to-end authentication, in contrast, allows user to verify the identity claimed by other users directly, without reference to a common trusted third party. For example, user software may run an authentication protocol end-to-end, to verify the identity claims of other users, and subsequently protect the authenticity or other properties of the communications cryptographically. Mutual authentication may be based on users initially sharing a short secret [32]; or they it rely on users verifying each other's public keys directly or indirectly [4].

The advantage of the end-to-end authentication approach is that it allows secure channels to be boot-strapped between users, and those channels are not subject to compromise by the trusted third party server in the first architecture (see Section 4.3.2 on channel security). Thus state of the art services should provide means, or at least support users that wish to authenticate others in an end-to end fashion.

### 4.1.3    Privacy in federated identity management and single-sign-on

Federated identity management systems separate entities that enrol users, and are able to identify them, from entities that rely on the result of the authentication process. In such systems a user may be registered as a user of a service A with a certain identity, and authenticate using this identity to a third party service B. In fact service B may allow authentication using any number of services, and accept the identities provided by those services.

A number of widely deployed single-sign-on (SSO) services are provided by major internet companies, with varying privacy properties. One federated identity management single-sign-on system with advanced privacy features is Shibboleth[30] used by a network of major United States educational institutions. Shibboleth is designed to allow users to provide only the necessary subset of their attributes to a service provider after successful authentication. Even the user-id is considered to be an attribute that may or may not be released depending on whether a service absolutely needs to uniquely identify users. Some use cases, e.g. an on-line library providing material to members of the university, may not need to know the exact user but merely their university membership status. This embeds and supports the data protection minimisation principle.

The policy implemented by Shibboleth, allowing selective disclosure of attributes, including partial anonymity, should be considered state of the art for federated identity systems. A number of similar designs such as InfoCards [9] and the Liberty Alliance protocols [10] have been proposed offering similar privacy protections, attesting to the maturity of those techniques. However, it is important to note that the privacy protections offered by all these protocols are only robust against the third party service observations. An eavesdropper looking at user traffic may be able to infer who users are, and their attributes; similarly the identity provider is made aware of every authentication session a user partakes in and the services they are accessing. Protocols with entities (including identity providers) that are able to observe the details of all authentication sessions of users should generally be avoided on the grounds they may be used or abused for pervasive monitoring. In Section 4.2 we discuss modern cryptographic mechanisms, namely selective disclosure credentials, allowing for selective revelation of attributes that is robust against such threats. Where possible, adopting techniques shielding users from network observers as well as their identity providers should be preferred.

## 4.2   Attribute based credentials

Attribute based credentials[31] are a fundamentally different technique to implement identity management, compared to the more traditional forms of federated identity management that are described in section 4.1.3

In federated (or network based) identity management, information about the identity of a user accessing a service is always obtained through an online identity provider (IdP). When accessing the service the user is redirected to this identity provider. Only after the user successfully logs in to this identity provider, this IdP relays the requested information about the user back to the service provider. This makes the IdP the spider in the web for all identity-related transaction, which leads to several security, privacy and usability concerns [8]. For example, the identity provider can log all service providers the user does business with while the service provider may obtain more personal information than strictly necessary for offering the service. If the information requested by the service provider is used to grant access to a valuable resource (say, for example, your health records or financial administration), then the identity provider effectively has the keys to this information: it can access this information without your consent or knowledge.

---

[30] http://net.educause.edu/ir/library/pdf/eqm0442.pdf 37

[31] Attribute based credentials have historically been called anonymous credentials. This term is no longer used as credentials can contain both identifying and non-identifying attributes. If they contain identifying attributes, they are by definition not anonymous.

Attribute based credentials (ABC) on the other hand put the user central to all transactions related to its identity.

### 4.2.1 Principles

Central to ABCs is the concept of an attribute. An attribute is any property that describes some aspect about a natural person[32]. The following items are examples of attributes: your name, age, date of birth, colour of hair, diplomas, grades, subscriptions, event tickets, to name but a few. Some attributes are static (like date of birth), others are dynamic (like a subscription to a newspaper). Some are identifying (your name) while others are not (your age, if the anonymity set is large enough).

Attribute based credentials allow a user (in this context called a prover) to securely and privately prove ownership of an attribute to a service provider (in this context called a verifier). This is why attributes are stored in a secure container called a credential. Conceptually a credential is very similar to a certificate. A credential is issued (and signed) by a credential issuer that is trusted to provide valid values for the attributes in the credential. For example, the government is a trusted source for your age and nationality, whereas the bank is a trusted source for your credit score and your high school a trusted source for your graders and diplomas. Credentials typically also contain an expiration date, and are linked to the private key of the prover. This prevents credentials from being used by others.

Credentials are fundamentally different from certificates in that they are never shown to other parties as is, as this would make the prover traceable. In fact, attribute based credential systems implement a so called selective disclosure protocol that allow the prover to select a subset of the attributes in a credential to be disclosed to the prover. The other attributes remain hidden. Certain schemes even allow the prover to only disclose the value of a function $f(a1,...,an)$ over some attributes $a_1,...,a_n$. Given your name, date of birth and place of birth, this allows you to prove you are over 18, without revealing anything more. Note that in general a prover may choose to reveal attributes that are in different credentials.

### 4.2.2 Properties

Attribute based credentials must satisfy the following properties (cf. [59, 49]).

**Unforgeability.** Only issuers can create valid credentials. The attributes contained within a credential cannot be forged. A selective disclosure can only involve attributes from valid (unrevoked, see below) credentials.

**Unlinkability**. Whenever a credential is used, no-one (except the prover or the inspector, see below) can link this to the case when this credential was issued. Moreover, whenever a credential is used, no-one (except the prover or the inspector) can link this credential to a case where it was used before. This implies that the issuer cannot determine where you use your credentials. Moreover, even if you use the same credential a hundred times at the same service provider, the service provider is completely unaware of this fact: as far as he is concerned he has served a hundred different customers. Note that revoked credentials (see below) are linkable by definition.

---

[32] Strictly speaking ABCs can also be used to describe properties of arbitrary entities, but we will not consider this for the purpose of this exposition.

**Non-transferability**. A credential issued to one prover cannot be used by another prover to prove possession of certain attributes. In particular, a single prover cannot try to combine several credentials that belong to different persons in a single selective disclosure. In other words users cannot collude and pool credentials to prove ownership of a combination of attributes not all owned by a single one of them.

**Revocability.** Credentials can be revoked. Once a credential is revoked, no verifier will accept disclosed attributes that involve this credential. (If a single credential is revoked, use of that credential is therefore linkable.)

**Identity escrow and black listing.** A trusted inspector has the power to retrieve the identity of the prover, when given a full trace of a selective disclosure between this prover and a verifier. Alternatively, a misbehaving user can be blacklisted in an anonymous way, cf [188]

The unlinkability property sets credentials apart from certificates. We note that not all systems implement identity escrow. Finally, revocation is actually hard to implement efficiently because it badly interacts with the unlinkability requirement. We refer to Lapon et al. [136] for a survey.

### 4.2.3    Basic techniques

Fundamentally, there are two basic strategies to implement unlinkability for attribute based credentials.

The first strategy requires provers to use credentials only once. Provers must request a fresh credential with the same set of attributes from the issuer each time they want to disclose an attribute from it. (Provers are allowed to cache credentials and request a batch of them whenever they are online.) This trivially implements unlinkability between multiple selective disclosures. Unlinkability of a credential between its issuing and the later disclosing of attributes in it is achieved by a blind signature protocol [58] that hides the actual credential from the issuer. U-Prove [42], one of the main attribute based credential schemes currently owned by Microsoft, is based on this principle. The main advantage of this scheme is that it allows efficient implementations. The main disadvantage is that provers need to be online and obtain fresh credentials all the time. Another state of the art protocol to issue such credentials is by Baldimtsi-Lysyanskaya [24].

The alternative strategy allows a prover to use a credential multiple times. In this case the prover must hide the actual credential from the user to prevent linkability. For this the powerful concept of a zero-knowledge proof is used. Very roughly speaking, zero knowledge proofs allow you to prove possession of a secret without actually revealing it. Moreover, the verifier of such a proof cannot convince anybody else of this fact. (Hence a traditional challenge response protocol is not zero knowledge.) Idemix [49], the other main attribute based credential scheme owned by IBM, is based on this principle. In Idemix, provers prove in zero knowledge that they own credentials, signed by the issuers, that contain the disclosed attributes. The advantages and disadvantages are exactly opposite to those of U-Prove: Idemix is more complex and thus has less efficient implementations. However, provers can be offline and use credentials multiple times without becoming linkable.

We do note however that recent research [202] has shown that efficient implementations of Idemix, that allow it to be used even on a smart card, are possible[33].

---

[33] http://www.irmacard.org

An efficient implementation of a multiple-show attribute based credential scheme does not really exist yet. However, it is to expect that in the near future hardware support for pairings becomes available. Furthermore, a promising (but so far only theoretical) approach appears to be the use of self-blindable credentials [198, 28]. Instead of proving ownership of a credential using a zero-knowledge protocol, provers randomise (in other words blind) the credential themselves before sending it to the verifier. This randomisation prevents linkability. Of course the randomisation procedure must maintain the values of the attributes and the possibility for the verifier to verify the issuer signature on it[34]

## 4.3 Secure private communications

Most physical network links provide poor guarantees of confidentiality and privacy. Local networks are increasingly wireless, and wide area networks are impossible to physically secure against pervasive surveillance. Therefore, any information from a user to a service or between users should preferably be encrypted using modern cryptographic techniques to render it unintelligible to eavesdroppers. All types of communications from the user should be protected: personal information or sensitive user input should be encrypted to preserve its privacy (and security); however, even accesses to otherwise public resources should be obscured through encryption to prevent an eavesdropper from inferring users' patterns of browsing, profiling, service use [97] or extracting identifiers that may be used for future tracking.

### 4.3.1 Basic Encryption: client-service

Deployed cryptographic channels provide a high degree of confidentiality for communications between clients and services when implemented and configured correctly. Widely deployed technologies include the latest Transport Layer Security 1.2 (TLS1.2) [79], as well as the Secure Shell (SSH) protocols [25]. Those technologies provide a confidential, and possibly authenticated channel between clients and end-services.

Both TLS and SSH use "public key cryptography" technologies that allow for encrypted channels to be set up without clients and servers sharing any prior secrets. However, TLS1.2 does rely on a public key certificate infrastructure of certificate authorities that is necessary to ensure the authenticity of the server end of the encrypted tunnels. Conversely, compromised authorities may lead to the security of channels being compromised. For public resources, accessible through web-browsers, a valid public key certificate is necessary and should be obtained. In case TLS is used within custom built software or mobile application, certificate pinning techniques may be used to avoid the need for a certificate from a globally trusted authority (Certificate Pinning [96]). Such a model should be preferred for internal channels such as those used by application for software updates. SSH relies on manual user verification of the service key, which is more taxing on non-technical users. Such a model should only be deployed after user studies confirm users are capable of performing checks consistently and correctly.

Deploying TLS 1.2, or an equivalent secure channel, for every network interaction should be considered the current state of the-art. Detailed guidance about the currently recommended ciphersuites for both families of protocols is provided in the ENISA report on ciphers and key lengths[180].

A number of technologies may be used for communication within an organisation to protect user data in transit over networks. One such example is IPSec [92] that creates secure communication tunnels between networked machines, or between networks connected by public network links. It is recommended that traffic internal to an organisation (local-area network) is encrypted if it may contain user information, such as for performing back-ups or communications between application and database servers. Links over wide area networks, be it public or private (lease fiber / line) should always be encrypted due to the difficulty of guaranteeing the physical security of those channels. A number of mature Virtual Private Network (VPN) technologies and products are commercially available to protect such links.

---

[34] This discussion sidesteps the issue of selectively disclosing attributes in a credential. Moreover, one can argue that Idemix is also based on the self-blinding principle as it randomises the signature on the credential before sending it to the verifier. Idemix can however not randomise all parts of the signature and hence needs to prove (in zero-knowledge) possession of an import component of that signature.

### 4.3.2 Basic Encryption: end-to-end

A number of services, such as Voice-over-Internet-Protocol (VoIP), electronic mail or instant messaging and social networking mediate communications between end users. Such services should prefer to encrypt the communications between users in an end-to-end fashion, meaning the encryption is added at one user end-point and is only stripped at the other end-user end-point, making the content of communications unintelligible to any third parties including the service providers. While the service providers may wish to assist users in authenticating themselves to each other for the purpose of establishing such an end-to-end encrypted channel, it is preferable, from a privacy perspective, that the keys used to subsequently protect the confidentiality and integrity of data never be available to the service providers, but derived on the end-user devices.

A number of services may require some visibility into the contents of communications for either routing them to the correct destination, or providing value added services. In such cases the minimum amount of information should be exposed to the provider, and to the extent possible any manipulation of the content of communication should be performed on user devices and minimising the content leaked to the service providers.

A number of technologies have been proposed, implemented and standardised to different extents to provide end-to-end confidentiality. The Pretty Good Privacy (PGP) [210] software as well as S/MIME standards [161] may be used to protect email correspondences end-to-end; the Off-The-Record messaging (OTR) [38] protocols are widely supported by clients software to protect instant messaging conversations. Mobile applications such as Whisper Systems' TextSecure provide end-to-end encrypted mobile chat; Crypto Phone and Red Phone provide end-to-end encrypted communications. As such the technology for providing end-to-end communications is mature, and provision of this property should be considered the state of the art.

### 4.3.3 Key rotation, forward secrecy and coercion resistance

Modern encryption technologies provide strong confidentiality guarantees, but only against eavesdroppers that are not able to get access to the secrets keys used to protect them. Those short binary strings therefore become valuable targets for theft, extortion, compromise through hacking or coercion.

To minimise the potential for keys to be compromised it is important to rotate keys regularly, therefore minimising the exposure of private data in case of a key compromise. Modern cryptographic systems, including some configurations of TLS (using the DH or ECDH cipher suites) and all configurations of OTR, provide forward secrecy, a property that guarantees the use of fresh keys for each communication session and discarding of key upon session termination. Forward secrecy guarantees that after a communication session is over, the secret key material cannot be recovered, and therefore no amount of coercion can render the encrypted material intelligible. The provision of such automatic key rotation schemes should be considered.

The state of the art is for no long-term secrets to be used to protect the confidentiality of interactive end-to-end conversations. For asynchronous communications (such as email) regular key rotation should be enforced and automatically managed by state-of-the-art services and applications.

## 4.4 Communications anonymity and pseudonymity

End-to-end encryption may be used to protect the content of communications, but leaves meta-data exposed to third-parties. Meta-data is information "about" the communication, such as who is talking to whom, the time and volume of messages, the duration of sessions or calls, the location and possibly identity of the network end-points.

The exposure of meta-data may have a devastating impact on privacy. Uncovering the fact that a journalist is talking to someone within an organisation or government department may compromise them as a journalistic source, even if the details of the message contents are not recoverable. Similarly, observing someone persistently browsing for information on some form of cancer may be indicative of a health concern or condition. Meta-data may also uncover lifestyle information that is not immediately obvious to communicating parties. For example persistent collocation of two mobile devices at out of office hours and on weekends is indicative of a close personal relationship. Meta-data analysis of mobile phone location logs, or WiFi / IP addresses, can uncover those relations even when the individuals concerned have not exchanged any messages.

### 4.4.1 Properties of anonymous communications

The family of privacy systems that obscure communications meta-data are known under the umbrella term "anonymous communications" [61][35]. However, those systems may provide a number of subtle privacy properties besides straight-forward anonymity. A system provides sender anonymity or initiator anonymity if it hides who sent a message, or initiated a network connection. Conversely, a system provides recipient anonymity if it is possible to contact the individual or service, without learning its physical or network location. The first property allows users to access services anonymously. The latter property is useful to run services without exposing the meta-data of the host or operators.

Recipient and initiator anonymity properties ensure that identifying information is hidden from both third parties, and their communication partners. A weaker, but very useful property, is third-party anonymity that ensures meta-data is not revealed to third parties, while both partners know, with high certainty, each other's identities. Third-party anonymity is closest to traditional secure channels, and simply augments the protection they provide by hiding meta-data. Variants of third-party anonymity include a number of pseudonymous communication channels: such systems hide the identity of either the initiator or responder, but assign to users a stable pseudonym that can be used to fight spam or abuse. It is important to note that long-term pseudonyms run the risk of becoming as revealing as real identities, as users perform linkable actions and leak an increasing amount of identifying or personal information over time. It is therefore good practice to allow users to refresh their pseudonyms or have control of more than one at a time.

Finally, specialised anonymous communications systems may require a high degree of robustness and strong guarantees that all messages are relayed correctly, without genuine messages dropped or adversary messages being injected. Such systems are used to provide anonymity for ballots of cryptographically protected elections, where dropping or injecting ballots could be catastrophic [127]. They thus have to provide cryptographic proofs of correct anonymisations and operation—these proofs are expected to be publicly verifiable.

---

[35] Note that the term "anonymous" in European law usually means that re-identification is not possible (at least not with proportionate effort) while the technological research typically refers to anonymity within a defined (or to be defined) anonymity set [150].

### 4.4.2    Threat models

Like any security system anonymous communications techniques provide properties subject to a threat model—i.e. a set of capabilities available to the adversary. Systems have been designed to protect users from adversaries able to observe parts of the network (partial adversary) or the whole network (global passive adversary). Since anonymous channels rely on open services to provide their protections one may assume that some of those services are also controlled by the adversary. Finally, the adversary may not only be able to observe the network or some services, but also tamper with the communications they handle to inject, delay, modify or drop messages (active adversary). When deploying anonymity system it is important to always evaluate whether the protections they provide withstand a sufficiently strong adversary for the purposes they are fielded.

### 4.4.3    Families of anonymous communications

As a basic principle one cannot hide their identity by themselves, so all systems providing anonymity properties attempt to conflate messages from multiple users in a way that is difficult to disentangle for the adversary. As a result anonymous communications systems benefit from large volumes of users, a phenomenon known as "anonymity loves company" [80]. As a result there might be benefits in a number of organisations running such systems collaboratively increasing the privacy of all their users.

**Single Proxies & VPNs.** The simplest means to achieve some degree of anonymity is to use a "proxies". Those proxies may take the form of open SMTP relays for e-mail, or open SOCKs proxies for generic TCP streams. However, anyone observing the traffic around the proxy can deduce the identity of those communicating. A slightly more robust mechanism involves the use of Virtual Private Networks (VPN) that are commercially available. Such services encrypt all traffic from the user to a gateway, and then allow it to exit to the open internet through a single (or few) points. Thus an adversary observing the client cannot simply deduce the destination of the traffic. However, a global passive adversary may be able to trace communications if their timing and volumes are not carefully hidden (which is uncommon due to costs).

**Onion Routing.** Single relays depend on a single operator that may be observed or coerced by an adversary to reveal the identities of communicating parties. The onion routing [108] family of schemes alleviates this by relying on multiple relays that possibly carry communications. As a result an adversary would have to control a much larger number of relays, or coerce a number of them to effectively track activities. The currently popular Tor service [81] has over 5000 such relays and serves over 1 million users per day. The Jondonym service[36] similarly operates a number of cascades of relays (although a smaller number) and provides a commercial service. Despite the increased difficulty, a global passive adversary may still be able to infer who is talking to whom through statistical analysis.

**Mix-networks.** Mix networks [61], in addition to using multiple relays restrict the size of messages to a uniform quantum, and are engineered to allow for long delays and cover traffic to mask statistical leaks that could trace messages. Such operational systems include mixmaster [146], and experimental systems providing support for replies have also been engineered and operated, such as mixminion [73]. The increase in latency over onion routing and the lack of bi-directional communications restricts their appeal, drives the number of users down, and as a result they may provide a lesser degree of anonymity despite their stronger designs. Special mix-networks designed for electronic and crypto-graphic elections are also available to anonymise ballots, while providing cryptographic proofs of correct operation. Such systems include the latest Verificatum[37].

---

[36] https://anonymous-proxy-servers.net/

**Broadcast schemes.** Finally, a simple but expensive method to achieve anonymity is broadcast with implicit addressing. Such schemes simply send all messages to everyone in a group without any designation of the recipient. Each receiver has to attempt to decode the message with their keys, to determine if it is destined to them. These schemes, as well as the DC-net scheme for sending anonymously, become expensive as groups grow, since every actual message requires all members of a group to send or receive a message. If however a physical network layer allows for cheap broadcast they are very competitive [184].

While some anonymous communications systems are now deployed, and their infrastructure is mature (as for Tor), they suffer from some fundamental limitations. Low-latency anonymity systems, such as onion routing, are not resistant to a network adversary that can observe traffic from the source and to the destination of a circuit [209]. Such an adversary is able, over a short time, to de-anonymise the connection. Higher latency anonymity systems may not suffer from such a short term attack, but suffer from long-term disclosure attacks [72], where an adversary can discover the long-term communication partners of senders and receivers. Their poor network performance makes them impractical for interactive application, and they struggle to support forward secure encryption mechanisms opening them to coercion attacks. Thus, anonymity properties supported by privacy-enhancing technologies, are inherently more costly and more fragile than equivalent secure channels and private authentication mechanisms.

### 4.4.4 Steganography and blocking resistance

In a number of settings other meta-data, besides the identity of conversing parties, needs to be protected to preserve privacy or related properties. In the context of "anonymous communications" it may be necessary to masquerade the type of service used, to bypass restrictions on communications put in place by service providers. The Tor project deploys an architecture of "Bridges" and "Pluggable Transports" [15] that aim to hide traffic to this service by making it look like an innocuous encrypted service.

The attempt to make Tor traffic look like other types of traffic is a specific application of a broad family of "steganographic" techniques [130]. Steganography is the discipline of how to modulate messages to make them look like a cover, in such a way that the fact that a hidden, or encrypted, communication has taken place is not detectable. Conversely, steganalysis is the discipline of analysing cover messages trying to detect whether, or not, they may contain a steganographic hidden message.

Some aspects of steganography are related to engineering privacy by design, as they can be used to implement coercion resistant properties. For example the disk encryption utility TrueCrypt [101] provides options for steganographic containers that allow users to encrypt files in a deniable manner. A mature research field exists on how to embed hidden messages in other media, such as images, audio or video, but few established products support such features for direct messaging by users.

## 4.5 Privacy in databases

The meaning of database privacy is largely dependent on the context where this concept is being used. In official statistics, it normally refers to the privacy of the respondents to which the database records correspond. In co-operative market analysis, it is understood as keeping private the databases owned by the various collaborating corporations. In healthcare, both of the above requirements may be implicit: patients must keep their privacy and the medical records should not be transferred from a hospital to, say, an insurance company. In the context of interactively queryable databases and, in particular, Internet search engines, the most rapidly growing concern is the privacy of the queries submitted by users (especially after scandals like the August 2006 disclosure by the AOL[38] search engine of 36 million queries made by 657000 users). Thus, what makes the difference is whose privacy is being sought.

---

[37] http://www.verificatum.org/

The last remark motivates splitting database privacy in the following three dimensions:

1.  *Respondent privacy* is about preventing re-identification of the respondents (e.g. individuals like patients or organisations like enterprises) to which the records of a database correspond. Usually respondent privacy becomes an issue only when the database is to be made available by the data collector (hospital or national statistical office) to third parties, like researchers or the public at large.
2.  *Owner privacy* is about two or more autonomous entities being able to compute queries across their databases in such a way that only the results of the query are revealed.
3.  *User privacy* is about guaranteeing the privacy of queries to interactive databases, in order to prevent user profiling and re-identification.

The technologies to deal with the above three privacy dimensions have evolved in a fairly independent way within research communities with surprisingly little interaction.

## 4.6 Technologies for respondent privacy: statistical disclosure control

Respondent privacy has been mainly pursued by statisticians and some computer scientists working in statistical disclosure control (SDC), also known as statistical disclosure limitation (SDL) or inference control. The seminal paper of SDC was written by Dalenius in 1974 [71]. A current state of the art can be found in the book [122].

Two types of disclosure are addressed by SDC. On the one hand, attribute disclosure occurs when the value of a confidential attribute of an individual can be determined more accurately with access to the released data than without. On the other hand identity disclosure occurs when a record in the anonymised data set can be linked with a respondent's identity. In general, attribute disclosure does not imply identity disclosure, and conversely.

### 4.6.1 Tabular data protection

There are several types of tables:

*   *Frequency tables.* They display the count of respondents (natural number) at the crossing of the categorical attributes. E.g. number of patients per disease and municipality.
*   *Magnitude tables.* They display information on a numerical attribute (real number) at the crossing of the categorical attributes. E.g. average age of patients per disease and municipality.
*   *Linked tables.* Two tables are linked if they share some of the crossed categorical attributes, e.g. "Disease" × "Town" and "Disease" × "Gender".

Whatever the type of the table, marginal row and column totals must be preserved. Even if tables display aggregate information, disclosure can occur:

---

[38] http://techcrunch.com/2006/08/06/aol-proudly-releases-massive-amounts-of-user-search-data/

- *External attack.* E.g., let a released frequency table "Ethnicity" × "Town" contain a single respondent for ethnicity $E_i$ and town $T_j$. Then if a magnitude table is released with the average blood pressure for each ethnicity and each town, the exact blood pressure of the only respondent with ethnicity $E_i$ in town $T_j$ is publicly disclosed.
- *Internal attack.* If there are only two respondents for ethnicity $E_i$ and town $T_j$, the blood pressure of each of them is disclosed to the other.
- *Dominance attack.* If one (or few) respondents dominate in the contribution to a cell in a magnitude table, the dominant respondent(s) can upper-bound the contributions of the rest. E.g. if the table displays the cumulative earnings for each job type and town, and one individual contributes 90% of a certain cell value, s/he knows her/his colleagues in the town are not doing very well.

SDC principles for table protection can be classified as follows.

- *Non-perturbative.* They do not modify the values in the cells, but they may suppress or recode them. Best known methods: cell suppression (CS), recoding of categorical attributes.
- *Perturbative.* They modify the values in the cells. Best known methods: controlled rounding (CR) and the recent controlled tabular adjustment (CTA).

**Cell suppression.** In this approach, sensitive cells are identified in a table, using a so-called sensitivity rule. Then the values of sensitive cells are suppressed (primary suppressions). After that, additional cells are suppressed (secondary suppressions) to prevent recovery of primary suppressions from row and/or column marginals. Examples of sensitivity rules for primary suppressions are:

*(n, k)*-**dominance**. A cell is sensitive if n or fewer respondents contribute more than a fraction k of the cell value.

*pq*-**rule.** If respondents' contributions to the cell can be estimated within q percent before seeing the cell and within p percent after seeing the cell, the cell is sensitive.

*p%*-**rule**. Special case of the *pq*-rule with *q* = 100.

As to secondary suppressions, usually one attempts to minimise either the number of secondary suppressions or their pooled magnitude (complex optimisation problems). Optimisation methods are heuristic, based on mixed linear integer programming or networks flows (the latter for 2-D tables only). Implementations are available in the τ-Argus package [123].

Controlled rounding and controlled tabular adjustment. CR rounds values in the table to multiples of a rounding base (marginals may have to be rounded as well). CTA modifies the values in the table to prevent inference of sensitive cell values within a prescribed protection interval. CTA attempts to find the closest table to the original one that protects all sensitive cells. CTA optimisation is typically based on mixed linear integer programming and entails less information loss than CS.

### 4.6.2 Queryable database protection

There are two main SDC principles for queryable database protection:

- *Query perturbation.* Perturbation (noise addition) can be applied to the microdata records on which queries are computed (input perturbation) or to the query result after computing it on the original data (output perturbation).
- *Query restriction.* The database refuses to answer certain queries.

**Differential privacy for output perturbation.** As defined in [93], a randomised query function F gives $\varepsilon$-differential privacy if, for all data sets D1, D2 such that one can be obtained from the other by modifying a single record, and all S ⊂ Range(F), it holds that

Pr(F(D1) ∈ S) ≤ exp($\varepsilon$) × Pr(F(D2) ∈ S).

Usually F(D) = f (D) + Y (D), where f (D) is a user query to a database D and Y (D) is a random noise (typically Laplace with zero mean and Δ( f )/o $\varepsilon$, where Δ( f ) is the sensitivity of f and $\varepsilon$ is a privacy parameter (the larger, the less privacy)). Hence, differential privacy follows the output perturbation principle.

***Query restriction.*** This is the right approach if the user does require deterministically correct answers and these answers have to be exact (i.e. a number). Exact answers may be very disclosive, so it may be necessary to refuse answering certain queries at some stage. A common criterion to decide whether a query can be answered is query set size control: the answer to a query is refused if this query together with the previously answered ones isolates too small a set of records. The main problem of query restriction are: i) the computational burden to keep track of previous queries; ii) collusion attacks can circumvent the query limit.

**Tracker attacks.** Query set size control is justified by the existence of trackers, pointed out already in 1979 by Denning et al. [76]. A tracker is a sequence of queries to an on-line statistical database whose answers disclose the attribute values for a small subgroup of individual target records or even a single record. In [77] it was shown that building a tracker is feasible and quick for any subgroup of target records.

### 4.6.3 Microdata protection

A microdata file X with s respondents and t attributes is an s × t matrix where $X_{ij}$ is the value of attribute j for respondent i. Attributes can be numerical (e.g. age, blood pressure) or categorical (e.g. gender, job). Depending on their disclosure potential, attributes can be classified as:

- *Identifiers.* Attributes that unambiguously identify the respondent (e.g. passport no., social security no., name-surname, etc.).
- *Quasi-identifiers or key attributes.* They identify the respondent with some ambiguity, but their combination may lead to unambiguous identification (e.g. address, gender, age, telephone no., etc.).
- *Confidential outcome attributes.* They contain sensitive respondent information (e.g. salary, religion, diagnosis, etc.).
- *Non-confidential outcome attributes.* Other attributes which contain non-sensitive respondent info.

Identifiers are of course suppressed in anonymised data sets. Disclosure risk comes from quasi-identifiers (QIs), but these cannot be suppressed because they often have high analytical value. Indeed, QIs can be used to link anonymised records to external non-anonymous databases (with identifiers) that contain the same or similar QIs; this leads to re-identification. Hence, anonymisation procedures must deal with QIs.

There are two principles used in microdata protection, data masking and data synthesis:

- Masking generates a modified version X' of the original microdata set X, and it can be *perturbative masking* (X' of the original microdata set X) or *non-perturbative masking* (X' is obtained from X by partial suppressions or reduction of detail, yet the data in X' are still true).
- Synthesis is about generating synthetic (i.e. artificial) data X' that preserve some preselected properties of the original data X.

**Perturbative masking.** There are several principles for perturbative masking:

- *Noise addition.* This principle is only applicable to numerical microdata. The most popular method consists of adding to each record in the data set a noise vector drawn from a N(0,αΣ), with Σ being the covariance matrix of the original data. Means and correlations of original data can be preserved in the masked data by choosing the appropriate α. Additional linear transformations of the masked data can be made to ensure that the sample covariance matrix of the masked attributes is an unbiased estimator for Σ.

- *Microaggregation.* Microaggregation [87] partitions records in a data set into groups containing each at least $k$ records; then the average record of each group is published. Groups are formed by the criterion of maximum within-group similarity: the more similar the records in a group, the less information loss is incurred when replacing them by the average record. There exist microaggregation methods for numerical and also categorical microdata.

- *Data swapping.* Values of attributes are exchanged among individual records, so that low-order frequency counts or marginals are maintained. Although swapping was proposed for categorical attributes, its rank swapping variant is also applicable to numerical attributes. In the latter, values of each attribute are ranked in ascending order and each value is swapped with another ranked value randomly chosen within a restricted range (e.g. the ranks of two swapped values cannot differ by more than p% of the total number of records).

- *Post-randomisation.* The PRAM method [133] works on categorical attributes: each value of a confidential attribute is stochastically changed to a different value according to a prescribed Markov matrix.

**Differential privacy in noise-added data sets.** $\varepsilon$-differential privacy can also be viewed as a privacy requirement to be attained when adding noise to data sets (not just to query outputs). An $o$ $\varepsilon$-differentially private data set can be created by pooling the $o$ $\varepsilon$-private answers to queries for the content of each individual record.

**Non-perturbative masking.** Principles used for non-perturbative masking include:

- *Sampling.* Instead of publishing the original data file, only a sample of it is published. A low sampling fraction may suffice to anonymise categorical data (probability that a sample unique is also a population unique is low). For continuous data, sampling alone does not suffice.

- *Generalisation.* This principle is also known as coarsening or global recoding. For a categorical attribute, several categories are combined to form new (less specific) categories; for a numerical attribute, numerical values are replaced by intervals (discretisation).

- *Top/bottom coding.* Values above, resp. below, a certain threshold are lumped into a single top, resp. bottom, category.

- *Local suppression.* Certain values of individual attributes are suppressed in order to increase the set of records agreeing on a combination of quasi-identifier attributes. This principle can be combined with generalisation.

**$k$-Anonymity, generalisation and microaggregation.** As defined in [177], a data set is said to satisfy $k$-anonymity if each combination of values of the quasi-identifier attributes in it is shared by at least $k$ records. The principles originally proposed to attain $k$-anonymity were generalisation and local suppression on the quasi-identifiers. Later, [90] showed that microaggregation on the projection of records on their quasi-identifiers was also a valid approach.

**Synthetic microdata generation.** Rubin [173] proposed this principle, which consists of randomly generating data in such a way that some statistics or relationships of the original data are preserved. The advantage of synthetic data is that no respondent re-identification seems possible, because data are artificial. There are downsides, too. If a synthetic record matches by chance a respondent's attributes, re-identification is likely and the respondent will find little comfort in the data being synthetic. Data utility of synthetic microdata is limited to the statistics and relationships pre-selected at the outset: analyses on random subdomains are no longer preserved. Partially synthetic or hybrid data are more flexible.

#### 4.6.4    Evaluation of SDC methods

Evaluation is in terms of two conflicting goals:

- Minimise the data utility loss caused by the method.
- Minimise the extant disclosure risk in the anonymised data.

The best methods are those that optimise the trade-off between both goals.

**Evaluation of tabular SDC methods.** For cell suppression, utility loss can be measured as the number or the pooled magnitude of secondary suppressions. For CTA or CR, it can be measured as the sum of distances between true and perturbed cells. The above loss measures can be weighted by cell costs, if not all cells have the same importance. As to disclosure risk, it is normally evaluated by computing *feasibility intervals* for the *sensitive cells* (via linear programming constrained by the marginals). The table is said to be safe if the feasibility interval for any sensitive cell contains the *protection interval* previously defined for that cell by the data protector.

**Evaluation of queryable database SDC methods.** For query perturbation, the difference between the true query response and the perturbed query response is a measure of utility loss; this can be characterised in terms of the mean and variance of the added noise (ideally the mean should be zero and the variance small). For query restriction, utility loss can be measured as the number of refused queries. Regarding disclosure risk, if query perturbation satisfies epsilon-differential privacy, the disclosure risk is proportional to parameter epsilon.

**Evaluation of micro data SDC methods.** Utility loss can be evaluated using either data use-specific loss measures or generic loss measures. The former measure to what extent anonymisation affects the output of a particular analysis. Very often, the data protector has no clue as to what the users will do with the anonymised data; in this case, generic utility loss measures can be used that measure the impact of anonymisation on a collection of basic statistics (means, covariances, etc., see [89]) or that rely on some score (such as propensity scores [205]).

To measure disclosure risk in micro data anonymisation, two approaches exist: *a priori* and *a posteriori*. The a priori approach is based on some privacy model (ε-differential privacy, *k*-anonymity, *l*-diversity, *t*-closeness, etc.) that guarantees an upper bound on the disclosure risk by design. The *a posteriori* approach consists of running an anonymisation method and then measuring the risk of disclosure (e.g. by attempting record linkage between the original and the anonymised data sets, by using analytical risk measures adapted to the anonymisation method, etc.).

#### 4.6.5    Main privacy models used for a priori privacy guarantees

We have mentioned above ε-differential privacy [93] and *k*-anonymity [177]. A number of *k*-anonymity extensions exists to address the basic problem of *k*-anonymity: while it is able to prevent identity disclosure (a record in a *k*-anonymised data set cannot be mapped back to the corresponding record in the original data set) in general it may fail to protect against attribute disclosure. To illustrate this, consider a *k*-anonymised medical data set in which there is a *k*-anonymous group (a group of *k* patient records sharing the quasi-identifier values) whose records have the same (or very similar) values for the confidential attribute "Disease" (e.g. "Disease" is AIDS for all of them). In this case, an intruder that manages to determine that the anonymised record for her target individual belongs to that *k*-anonymous group learns that the target individual suffers from AIDS (even if that individual is remains *k*-anonymous). *k*-Anonymity extensions include the following:

- *l-Diversity* [142]. A *k*-anonymous data set is said to satisfy it if, for each group of records sharing quasi-identifier values, there are at least *l* "well-represented" values for each confidential attribute.

- *t-Closeness* [139]. A *k*-anonymous data set is said to satisfy it if, for each group of records sharing quasi-identifier values, the distance between the distribution of each confidential attribute within the group and the distribution of the attribute in the whole data set is no more than a threshold *t*.

See [91] for a critical assessment of *k*-anonymity and the above extensions. In general, they provide less strict privacy guarantees than $\varepsilon$-differential privacy, even if *t*-closeness with a suitable distance can yield $\varepsilon$-differential privacy [182]. However, *k*-anonymity-like models usually entail less utility loss than $\varepsilon$-differential privacy, which is attractive when one wishes to publish anonymised data sets to be used by researchers; see [65] for a comparison of both model families.

### 4.6.6 Software for a priori and a posteriori anonymisation

Regarding a priori anonymisation, software offering anonymisation with k-anonymity, l-diversity and t-closeness guarantees is freely available, e.g. ARX[39].

Also, a number of academic systems implement general purpose data processing systems with differential privacy guarantees. The Microsoft Research PINQ project[40] defines queries in a LINQ-like syntax, executes them and protects the results using a differentially private mechanism. The Airavat system[41] uses the Sample-and-aggregate differentially private mechanism to protect map-reduce query results using differential privacy. Other systems include Fuzz (UPenn) and GUPT (Berkeley).

In what respects a posteriori anonymisation, the official statistics community has produced the free software packages for microdata protection: $\mu$-ARGUS [124] and sdcMicro [186].

### 4.6.7 De-anonymisation attacks

For datasets with low to moderate dimension (number of attributes), anonymisation methods can protect against re-identification based on crossing databases. For high-dimensional datasets, protection becomes much more difficult. In [149], a new class of statistical de-anonymisation attacks against high-dimensional data sets was presented and illustrated on the Netflix Prize data. This data set contains anonymous movie ratings of 500,000 subscribers of Netflix and the attack showed that knowing a little bit about an individual subscriber is enough to find his record in the data set. In such high-dimensional data sets, it is hard to split attributes between quasi-identifiers and confidential, so *k*-anonymity-like methods are hardly usable. $\varepsilon$-differential privacy could still be used to prevent re-identification, but at the cost of a huge utility loss.

## 4.7 Technologies for owner privacy: privacy-preserving data mining

Owner privacy is the primary though not the only goal of privacy-preserving data mining (PPDM). PPDM has become increasingly popular because it allows sharing sensitive data for analysis purposes. It consists of techniques for modifying the original data in such a way that the private data and/or knowledge (i.e. the rules) remain private even after the mining process [199]. PPDM may provide also respondent privacy as a by-product.

PPDM based on random perturbation was introduced by [6] in the database community. This type of PPDM is largely based on statistical disclosure control (see Section 4.6 above). Independently and simultaneously, PPDM based on secure multiparty computation (MPC) was introduced by [140] in the cryptographic community. We next discuss some specifics of PPDM for data hiding and for knowledge hiding.

---

[39] http://arx.deidentifier.org/
[40] http://research.microsoft.com/en-us/projects/pinq/
[41] http://z.cs.utexas.edu/users/osa/airavat/

### 4.7.1    PPDM for data hiding

The approach can be oriented to countering attribute disclosure (protecting values of confidential attributes) or identity disclosure (protecting against re-identification).

Random perturbation PPDM for data hiding uses the statistical disclosure control methods discussed in Section 4.6 above. Cryptographic PPDM for data hiding uses SMC for a wide range of clustering and other data mining algorithms in distributed environments, where the data are partitioned across multiple parties. Partitioning can be vertical (each party holds the projection of all records on a different subset of attributes), horizontal (each party holds a subset of the records, but each record contains all attributes) or mixed (each party holds a subset of the records projected on a different subset of attributes). For example, a secure scalar product protocol based on cryptographic primitives is applied in privacy preserving *k*-means clustering over a vertically distributed dataset in [194].

Using SMC protocols which are based on cryptography or sharing intermediate results often requires changing or adapting the data mining algorithms. Hence, each cryptographic PPDM protocol is designed for a specific data mining computation and in general it is not valid for other computations. In contrast, random perturbation PPDM is more flexible: a broad range of data mining computations can be performed on the same perturbed data, at the cost of some accuracy loss.

### 4.7.2    PPDM for knowledge hiding

Knowledge hiding or rule hiding [200] refers to the process of modifying the original database in such a way that certain sensitive (confidential) rules are no longer inferable, while preserving the data and the non-sensitive rules as much as possible. A classification rule is an expression r : X → C, where C is a class item (e.g. Credit=yes in a data set classifying customers between those who are granted credit or not) and X is an itemset (set of values of attributes) containing no class item (e.g. Gender=female, City=Barcelona). The support of a rule r : X → C is the number of records that contain X , whereas the confidence of the rule is the proportion of records that contain C among those containing X . Rule hiding techniques change the data to decrease the confidence or support of sensitive rules to less than the minimum confidence or support threshold required for a rule to be inferred. Data perturbation, data reconstruction, data reduction and data blocking are some principles that have been proposed to implement rule hiding.

## 4.8    Technologies for user privacy: private information retrieval

User privacy has found solutions mainly in the cryptographic community, where the notion of private information retrieval was invented (PIR, [63]). In PIR, a user wants to retrieve an item from a database or search engine without the latter learning which item the user is retrieving. In the PIR literature the database is usually modelled as a vector. The user wishes to retrieve the value of the *i*-th component of the vector while keeping the index *i* hidden from the database. Thus, it is assumed that the user knows the physical address of the sought item, which might be too strong an assumption in many practical situations. Keyword PIR [64] is a more flexible form of PIR: the user can submit a query consisting of a keyword and no modification in the structure of the database is needed.

The PIR protocols in the above cryptographic sense have two fundamental shortcomings which hinder their practical deployment:

- The database is assumed to contain n items and PIR protocols attempt to guarantee maximum privacy, that is, maximum server uncertainty on the index *i* of the record retrieved by the user. Thus, the computational complexity of such PIR protocols is *O(n)*. Intuitively, all records in the database must be "touched"; otherwise, the server would be able to rule out some of the records when trying to discover *i*. For large databases, *O(n)* computational cost is unaffordable.

- It is assumed that the database server co-operates in the PIR protocol. However, it is the user who is interested in her own privacy, whereas the motivation for the database server is dubious. Actually, PIR is likely to be unattractive to most companies running queryable databases, as it limits their profiling ability. This probably explains why no real instances of PIR-enabled databases can be mentioned.

If one wishes to run PIR against a search engine, there is another shortcoming beyond the lack of server co-operation: the database cannot be modelled as a vector in which the user can be assumed to know the physical location of the keyword sought. Even keyword PIR does not really fit, as it still assumes a mapping between individual keywords and physical addresses (in fact, each keyword is used as an alias of a physical address). A search engine allowing only searches of individual keywords stored in this way would be much more limited than real engines like Google and Yahoo. In view of the above, several relaxations of PIR have been proposed. These fall into two categories:

- *Standalone relaxations.* In [88], a system named Goopir is proposed in which a user masks her target query by ORing its keywords with *k*−1 fake keyword sets of similar frequency and then submits the resulting masked query to a search engine or database (which does not need to be aware of Goopir, let alone co-operate with it). Then Goopir locally extracts the subset of query results relevant to the target query. Goopir does not provide strict PIR, because the database knows that the target query is one of the *k* OR-ed keyword sets in the masked query. TrackMeNot [121] is another practical system based on a different principle: rather than submitting a single masked query for each actual query like Goopir, a browser extension installed on the user's computer hides the user's actual queries in a cloud of automatic "ghost" queries submitted to popular search engines at different time intervals. Again, this is not strict PIR, because the target query is one of the submitted ones.
- *Multi-party relaxations.* The previous standalone approaches rely on fake queries and this can be viewed as a weakness: it is not easy to generate fake queries that look real and, besides, overloading the search engines/databases with fake queries clearly degrades performance. Multi-party relaxations avoid fake queries by allowing one user to use the help of other entities either for anonymous routing or for cross-submission:
    - An onion-routing system like Tor[42] is not intended to offer query profile privacy because it only provides anonymity at the transport level; however, if complemented with the Torbutton component, Tor can offer application-level security (e.g. by blocking cookies) and it can be used for anonymous query submission. Similarly, JonDonym [43] can provide application-level security with the recommended modified browser and add-ons.
    - In proxy-based approaches, the user sends her query to a proxy, who centralises and submits queries from several users to the database/search engine. Examples of proxies are DuckDuckGo[44], Ixquick[45], Startpage[46], Yippy[47], etc. While using the proxy prevents the database/search engine from profiling the user, the proxy itself can profile the user and this is a weakness.

---

[42] The Tor Project, 2014. http://www.torproject.org/.

[43] JonDonym, 2014. https://anonymous-proxy-servers.net/.

[44] DuckDuckGo, 2014. http://DuckDuckGo.com/.

[45] ixquick, 2014. http://www.ixquick.com/.

[46] Startpage, 2014. http://startpage.com/.

[47] yippy, 2014. http://yippy.com/.

   – Another principle is peer-to-peer: a user can submit queries originated by peers and conversely. In this way, peers use each other's queries as noise. The advantage here is that this noise consists of real queries by other peers, rather than fake queries as in standalone systems. The game-theoretic analysis in [86] shows that cross-submission among peers is a rational behaviour. Proposals exploiting this approach include Crowds [165] and more recently [85, 51, 201].

PIR techniques based on single or multiple servers are well understood scientifically, and library code with robust implementations is available (such as percy++[48]). Scalability issues however limit is applicability to databases of a few million records. Including those libraries into working products would involve R&D investments but should be considered an option when the number of records to be served make this technique theoretically possible.

## 4.9 Storage privacy

Storage privacy refers to the ability to store data without anyone being able to read (let alone manipulate) them, except the party having stored the data (called here the data owner) and whoever the data owner authorises. A major challenge to implement private storage is to prevent non-authorised parties from accessing the stored data. If the data owner stores data locally, then physical access control might help, but it is not sufficient if the computer equipment is connected to a network: a hacker might succeed in remotely accessing the stored data. If the data owner stores data in the cloud, then physical access control is not even feasible. Hence, technology-oriented countermeasures are needed, which we next review.

### 4.9.1 Operating system controls

User authentication and access control lists managed by the operating system are the most common way to guarantee some level of storage privacy. However, if an attacker gains physical access to the computer or is able to bypass the operating system controls, he can access the data.

### 4.9.2 Local encrypted storage

Locally storing the data in encrypted form is a straightforward option. One can use *full disk encryption* (FDE) or *file system-level encryption* (FSE).

In FDE, the entire content of the disk is encrypted, including the programs that encrypt bootable operating system partitions. Encryption can be done using disk encryption software or hardware.

In contrast, FSE encrypts the contents of files, but not the file system metadata, such as directory structure, file names, sizes or modification timestamps. FSE offer insufficient protection in case the metadata themselves need to be kept confidential.

A desirable property in encrypted storage is *forward secrecy*. This means that if a long-term key is discovered by an adversary, the keys derived from that long-term key to encrypt the data must not be compromised. Using a non-deterministic algorithm to derive data-encrypting keys from long-term keys is one way to achieve forward secrecy.

Examples of FSE systems include:

- FileVault, used in Mac computers; encryption and decryption are performed on the fly, in a way transparent to the user.

---

[48] http://percy.sourceforge.net/

- TrueCrypt is a source-available free ware utility (discontinued in May 2014) that was used for on-the-fly encryption. It can create a virtual encrypted disk within a file, encrypt a partition or the entire storage device.
- BitLocker is an FDE utility included with some versions of Windows. By default it uses the Advanced Encryption Standard in cipher-block chaining model. If a Trusted Platform Module (TPM) is available, it can be used by BitLocker to hold the disk encryption key and perform encryption/decryption. With a TPM, FDE is transparent to the user, who can power up and log into Windows as if there was no FDE.
- LUKS (Linux Unified Key Setup, [102]) is a disk encryption specification created for Linux. The reference implementation for LUKS operates on Linux using dm-crypt as the disk encryption software.

### 4.9.3 Format-preserving encryption

Format-preserving encryption encrypts a plaintext of some specified format into a ciphertext of identical format. Examples include encrypting a valid credit card number into a valid credit card number, encrypting an English word into an English word, etc.

One motivation for FPE is to integrate encryption into existing applications, with well-defined data models that establish the formats for the values of the various attributes in databases.

Conceptually, FPE can be viewed as a random permutation of the plaintext domain, because the ciphertext domain is exactly the plaintext domain.

However, for large domains, it is infeasible to pre-generate and remember a truly random permutation. Hence, the problem of FPE is to generate a pseudorandom permutation from a secret key, in such a way that the computation time for a single value is small.

Although the security of FPE was first formalized in [31], FPE constructions and algorithms were pre-existing, like [34,147].

### 4.9.4 Steganographic storage

Steganographic file systems [11] allow storing private information while hiding the very existence of such information. This mechanism gives the user a very high level of privacy: the user data are steganographically embedded in cover data.

A steganographic file system delivers a file to a user who knows its name and password; but an attacker who does not possess this information and cannot guess it, can gain no information about whether the file is present, even given complete access to all hardware and software. Hence, steganographic storage is especially attractive in a cloud setting, in which the hardware and the software are not under the user's control. Of course, in a cloud setting, the information embedding and recovery should be carried out by the user locally.

The downsides of steganographic storage are space inefficiency (a lot of cover data are needed to embed the actual user data in them), and possible data loss due to data collision or loss of the embedding key.

TrueCrypt supports steganographic storage by allowing hidden volumes: two or more passwords open different volumes in the same file, but only one of the volumes contains actual secret data.

### 4.9.5 Secure remote storage

Encrypted and steganographic storage can also be used to obtain private storage in remote settings, like cloud storage. In fact, steganographic storage is especially attractive in a cloud setting, in which the hardware and the software are not under the user's control.

In this case, however, the encryption/embedding and decrypt/recovery operations must be carried out locally, because the keys used in them must remain in the power of the user if any storage privacy is to be achieved.

Outsourced bulk data storage on remote "clouds" is practical and relatively safe, as long as only the data owner, not the cloud service holds the decryption keys. Such storage may be distributed for added robustness to failures. Such systems are available, as in the case of Tahoe-LAFS, which is commercially supported free software.[49]

A shortcoming of encrypted cloud storage compared with local encrypted storage is the leakage of access patterns to data: the cloud provider or a third partly may observer which blocks are accesses, modified and when. The pattern of activity may in itself leak private information. Techniques such as oblivious transfer hide such meta-data securely, but are not yet either scalable to large stores or flexible enough for general purpose storage. Approaches that attempt to obscure the pattern of accesses using naïve dummy queries have been repeatedly shown to be a weak privacy measure [190].

See Chapter 5 of [180] for more details on securing cloud storage.

### 4.9.6   Search on encrypted data

While storing encrypted data locally or remotely in cloud servers is a good way to achieve private storage, it implies sacrificing functionality to security and privacy. For example, if the user wishes to retrieve only documents containing certain keywords, one may be forced to decrypt all encrypted documents and search for the target keyword in the cleartext files. In a cloud setting, one must add the download time to the decryption time, so the task may be really cumbersome.

Being able to search on encrypted data is very attractive, especially in cloud systems, as the remote user can delegate the search to the cloud system, which normally has more computing power and holds the encrypted data locally.

In [181], the problem of searching on encrypted data was stated as follows. Alice has a set of documents and stores them on an untrusted server Bob. Because Bob is untrusted, Alice wishes to encrypt her documents and only store the ciphertext on Bob. Each document can be divided up into 'words'. Each 'word' may be any token; it may be a 64-bit block, a natural language word, a sentence, or some other atomic entity, depending on the application domain of interest. Assume Alice wants to retrieve only the documents containing the word W. A scheme is needed so that after performing certain computations over the ciphertext, Bob can determine with some probability whether each document contains the word W without learning anything else.

In [181], some practical schemes allowing this were described. After that, such schemes have come to be collectively known under the name *searchable encryption*, on which substantial literature exists. Nowadays, there are two types of searchable encryption.

- *Symmetric Searchable Encryption* (SSE) encrypts the database using a symmetric encryption algorithm. SSE has made much progress in recent years. For example [70] reports encrypted search on databases with over 13 million documents. Also, in [129], the first searchable symmetric encryption scheme was presented that satisfies the following important properties: sublinear search time, security against adaptive chosen-keyword attacks, compact indexes and the ability to add and delete files efficiently. More references on SSE can be found in [180].
- *Public-key Searchable Encryption* (PEKS) encrypts the database using a public-key encryption scheme. With PEKS, anyone (not just the data owner who encrypted the data) can search a certain keyword in the encrypted data. While this flexibility may be an advantage for some applications, it can also constitute a weakness. Details on PEKS can be found in [3].

---

[49] https://leastauthority.com

## 4.10 Privacy-preserving computations

### 4.10.1 Homomorphic encryption

Privacy homomorphisms (PH) were introduced in [167] as encryption transformations mapping a set of operations on cleartext to another set of operations on ciphertext. If addition is one of the ciphertext operations, then it was shown that a PH is insecure against a chosen-cleartext attack. Thus, a PH allowing full arithmetic on encrypted data is at best secure against known-cleartext attacks.

In fact, all examples of PHs in [167] were broken by ciphertext-only attacks or, at most, known-cleartext attacks [43]. Several homomorphic cryptosystems were also proposed that allow only one operation to be carried out on encrypted data (equivalent to either addition or multiplication on clear data). Such cryptosystems are called *partially homomorphic* and they include RSA [168], ElGamal [103], the well-known semantically secure Paillier cryptosystem [153] and several others.

In [83, 84], PHs allowing both addition and multiplication to be performed on encrypted data were presented; the PHs in those two papers could withstand ciphertext-only attacks, but it turned out they could be broken by a known-cleartext attack.

The breakthrough in homomorphic encryption came with the invention of Fully Homomorphic Encryption (FHE) by Gentry [106]. FHE is an encryption scheme that allows any arithmetic circuit to be applied to ciphertexts and obtain an output ciphertext that encrypts the output that would be obtained if the circuit was directly applied to cleartexts. Somewhat Homomorphic Encryption (SHE) are schemes that can evaluate only circuits of bounded multiplicative depth.

The state of the art of SHE schemes is that there are constructions that efficiently evaluate relatively simple circuits. The most efficient SHE-schemes are those based on a lattice problem called Ring-LWE. For example, the BGV scheme [41] and the scheme based on the NTRU cryptosystem [141]. Another family of efficient schemes is based on integer approximate-GCD problem [196]. Some of these efficient SHE constructions (e.g. [41]) support SIMD evaluation, which allows efficient processing of bulky data. See [180] for more references on homomorphic systems.

Computations relaying primarily on adding encrypted data items, and performing only a limited depth of multiplication on secrets, can be practically executed. Example protocols for computing simple statistics for smart metering deployments have been proposed and implemented [128], or even for training and evaluating simple regression models [109]. We recommend designers to consider SHE to compute simple but high-value results on private data. However, FHE or even SHE, cannot at this time be considered an alternative to running any program on cleartext data, due to its relatively low performance in general. This means that the privacy of data used for general computations on remote servers or clouds is inherently vulnerable against adversaries with local or logical access to those infrastructures.

### 4.10.2 Secure multi-party computation

Secure multi-party computation is also known as secure computation or multiparty computation (MPC). MPC methods enable several parties to jointly compute a function over their inputs, while at the same time keeping these inputs private. In the most general case, the joint computation may have one private output for each party.

MPC was first introduced by Andrew C. Yao [208] with the motivation of the "millionaire problem": two millionaires wish to compute which one is richer, but without revealing to each other how much money they have. MPC has been more general described in [63a,107]. Applications of MPC include privacy-preserving data mining (PPDM, e.g. [140]) and private information retrieval (PIR), mentioned in Section 4.7 above. E-voting systems (e.g. SCYTL [166], Prêt à Voter [175]) can also be regarded as a special case of MPC, in which the set of voters wish to compute the tally in such a way that the ballot of each voter stays private. It must be said, however, that one tends to use solutions for those applications that are less computationally involved than MPC.

Security in MPC can be passive or active. Passive security means that parties are honest-but-curious, that is, they follow the protocol but may try to learn the inputs of other parties. Active security assumes that the privacy of inputs (and maybe outputs) is preserved even if parties arbitrarily deviate from the protocol.

The following two are important primitives to build MPC protocols:

- *Oblivious transfer*. An oblivious transfer is a protocol in which a sender transfers one of potentially many pieces of information to a receiver, but remains oblivious as to what piece if any has been transferred. In [132] it was shown that oblivious transfer is complete for MPC: that is, given an implementation of oblivious transfer, it is possible to securely evaluate any polynomial-time computable function without any additional primitive.
- *Secret sharing and verifiable secret sharing*. Secret sharing [172, 35] refers to methods for distributing a secret amongst a group of participants, each of whom is allocated a share of the secret. The secret can be reconstructed only when a sufficient number, of possibly different types, of shares are combined together; the sets of shares which allow reconstructing the secret are collectively called an *access structure*. A set of shares that is not in the access structure is of no use. Verifiable secret sharing (VSS) schemes allow participants to be certain that no other players are lying about the contents of their shares, up to a reasonable probability of error. In [160] a VSS protocol was presented and it was shown that it could be used to implement any MPC protocol if a majority of the participants are honest. Furthermore the MPC thus achieved are information-theoretically secure: the privacy of the inputs is guaranteed without making any computational assumptions.

See [180] for additional references on MPC schemes and in particular for commercial systems implementing passively secure MPC.

Privacy-friendly computations based on MPC are practical for certain categories of simple but potentially high value scenarios. Secret sharing based MPC has been used to compute the outcome of beet auctions in Denmark[50] – a rather complex protocol. Secret sharing based MPC is also being trialled for computing privately aggregates of smart meter consumption in the Netherlands[74], and enables a product for protecting passwords[51]. There is good tool support for expressing and compiling computations into MPC, such as the free fairplay compiler [147], the commercial sharemind system[52]. As such for the use and deployment of solutions based on MPC, when applicable, should be encouraged.

## 4.11 Transparency-enhancing techniques

Several types of tools or functionalities have been proposed to enhance transparency and to place users in a better position to understand what data about them are collected and how they are used. Transparency-enhancing technologies (or: transparency tools) may comprise various properties that promote privacy and data protection [115, 118, 52]: (1) provision of information about the intended or actual privacy-relevant data processing and related risks, (2) overview of what privacy-relevant data the user has disclosed to which entity under which policies, (3) support of the user in "counter profiling" capabilities for extrapolating the analysis of privacy-relevant data, in particular with respect to group profiling. An additional functionality is the combination with online access to personal data and possibilities to exercise data subject rights such as withdrawing consent or requesting rectification, blocking and erasure of personal data (cf. Section 4.12).

---

[50] http://csrc.nist.gov/groups/ST/PEC2011/presentations2011/toft.pdf
[51] http://www.dyadicsec.com/technology/
[52] https://sharemind.cyber.ee/

Transparency-enhancing techniques cannot be realised by technological tools alone, but need to be intertwined with processes that provide the necessary information. Also, support of transparency for users has to be ensured no matter if the user employs specific technological solutions or not. In the following, the necessary processes for achieving transparency and complying with all of the notification obligations (of users or other data subjects as well as supervisory authorities, cf. Section 3.2) will not be further discussed, but they have to be considered in each specific case of privacy by design.

Since transparency in this context aims at individuals' understanding of data processing and related risks to provide a fair basis for informational self-determination, specific attention has been paid to usability [99] as well as accessibility and inclusion when designing transparency mechanisms and determining the ways to communicate information. Not all users are interested in all details of data processing, others would not be satisfied if they could only get high-level information. Here a common approach, also supported by the Art. 29 Data Protection Working Party [20], is to make available the information in multiple levels of detail, as needed by the individual.

Transparency-enhancing technologies can be classified into different categories depending on the underlying trust model:

- At one end of the spectrum, certain companies (such as Google) provide a privacy dashboard showing, for example, the type of personal data that they collect, how they are used, to what parties they are made visible[53], etc. Such dashboards have to be carefully designed to ensure that they do not mislead users and actually do not worsen rather than improve the situation [138]. Since the information is provided in a declarative mode, the subject has to trust the service provider for presenting the situation in a fair and comprehensive way.

- Other tools extract by themselves the privacy information rather than depending on the declarations of the service providers. For example, Lightbeam[54] (formerly Collusion) is a Firefox browser add-on that analyses the events occurring when a user visits a website and continuously updates a graph showing the tracking sites and their interactions. In the same spirit, tools such as TaintDroid [94] or Mobilitics [5] have been developed to detect information flows on smart phones (especially through third-party applications) and to report them to the users [95, 5]. A more comprehensive approach has been realised in the identity management projects PRIME and PrimeLife, the so-called Data Track [100, 54, 12], and further developed in the project A4Cloud [98]. This user-side function keeps track of the disclosure of personal data and logs which attributes (including pseudonyms) have been revealed to which entity under which conditions. This makes it easier for users to employ the same pseudonyms (or, if desired, not the same, of course) as in previous transactions with the same communication partner. The Data Track also provides an overview for later checks by the user (see also Section 4.12).

---

[53] In some cases, such as the Google dashboard, users can also manage their privacy settings and access (part of) their data through the dashboard.

[54] https://www.mozilla.org/fr/lightbeam/

- Another type of support to website users is offered by sites such as ToS;DR[55] (Terms of Service; Didn't Read) and TOSBack[56]. These sites rely on the effort of communities of peers (or experts) to evaluate privacy policies and track their evolution. This approach can be seen as a way to alleviate the inherent imbalance of powers between subjects and data controllers through a collaboration between subjects.
- Furthermore seals and logos might help to reduce the user burden, eg, the RFID logo[57]

Several languages are also available to make it easier for service providers and users to express their privacy policies and privacy requirements (sometimes called "privacy preferences") respectively. Declarations in these languages can be automatically translated into a machine-readable format. The privacy policy of a site can then be matched with the privacy requirements of the user and appropriate decisions taken (or information provided to the user) depending on the result of the matching.

The first and most well-known framework providing this kind of facilities are P3P[58] and Privacy Bird[59]. P3P allows website managers to declare their privacy policies and Privacy Bird displays different icons and acoustic alerts to inform the user and let her take the decision to visit the site, to have a closer look at its privacy policy, or to avoid it. Criticism has been raised against P3P through [164], both on the technical side and on the legal side[60]. Firstly, the categories of data that can be used to specify policies may be too coarse in many situations and, as a result, users are driven into excessive data disclosure. Secondly, a limitation of P3P is a lack of clarity which leaves the way open to divergent interpretations of privacy policies.

A possible way to address this issue is to endow the privacy policy language with a mathematical semantics. For example, CI [26] (Contextual Integrity) is a logical framework for expressing and reasoning about norms of transmission of personal information. CI makes it possible, for example, to express the fact that an agent acts in a given role and context. S4P [29] is an abstract[61], generic language based on notions of declaration and permission. A distinctive feature of S4P is the support of credential-based delegation of authority (allowing for example a user to delegate authority to perform certain actions to a certification authority). SIMPL [139, 144] is a more concrete language allowing users to express their policies using sentences such as "I consent to disclose my identity to colleagues when I meet them in the premises of my company". In contrast with P3P, this kind of sentence is associated with a formal semantics in a mathematical framework, which makes it possible to prove properties about the policies or to show that a given implementation is consistent with the semantics[62]. It should be stressed however that, even though they may have an impact in the design of future languages, these proposals are not currently supported by widely usable tools.

The channels to inform individuals are not limited to natural language and machine-readable items: privacy icons as well as other visual support are being proposed [100, 115, 69, 12], and even the acoustic signals of the Privacy Bird depending on the matching results between a given privacy policy and the own privacy requirements constitute a valuable way to help users' comprehension. It is not a trivial task to define and design appropriate icons or other means that are easily understandable by many users in a globalised world. When it comes to expressing specific privacy-enhancing functionality, e.g. concerning data minimising effects, even more challenges occur because the concepts are sometimes hard to grasp and may be counter-intuitive. Both privacy policy languages and icons or other audio-visual means could tremendously profit from standardisation.

---

[55] http://tosdr.org/

[56] https://tosback.org/

[57] http://europa.eu/rapid/press-release_IP-14-889_en.htm

[58] Platform for Privacy Preferences, http://www.w3.org/P3P/ [69]

[59] http://www.privacybird.org/

[60] The question of the validity of consent conveyed through this kind of tool is not straightforward [140, 157].

[61] S4P is abstract in the sense that, in order to keep the language simple, the choice is made to avoid specifying the semantics of service behaviours, leaving scope for the integration of specific or application dependent actions such as, for example, "informing third parties" or "deleting data".

[62] In other words, that the system behaves as expected by the user.

## 4.12 Intervenability-enhancing techniques

"Intervenability", as introduced in the explanation on privacy protection goals in Section 3.1, means the possibility to intervene and encompasses control by the user, but also control by responsible entities over contractors performing data processing on their behalf. Typical examples from the individual's perspective are giving, denying or withdrawing consent; exercising the rights to access (although this can also be regarded as a transparency functionality), to rectification, to blocking and to erasure of personal data; entering and terminating a contract; installing, de-installing, activating and de-activating a technical component; sending requests or filing complaints concerning privacy-related issues; involving data protection authorities or bringing an action at law.

It can be easily seen that intervenability techniques can often not be implemented solely on a technical level; instead, many processes of our democratic society and in particular of the juridical systems contribute to effective intervenability. Still, there are a few possibilities for technological support of the processes that have to be realised by the data processing entities, e.g. for enabling users to exercise their rights (cf. Section 2.2).

The Data Track, as already mentioned in Section 4.11, provides users online functionality for accessing their personal data (insofar this was offered by the communication partners) and helps them with requests concerning rectification and erasure [100, 115].

For privacy by design, it is essential to assist users and support their intervention possibilities. This begins already when users can easily conclude a contract and become a customer, but can hardly dissolve the contract and enforce the erasure of personal data that is not necessary anymore. For smart homes, intervenability could mean to be able to temporarily or permanently de-activate the sensors for those whose privacy-relevant data might be processed. This example shows the relation to social norms, e.g. whether and how guests of a smart home should also be able to express under which conditions they agree that their data is processed in the smart home or not.

## 5   Conclusions & Recommendations

This report is a first step towards establishing a set of guidelines for designing privacy-friendly and legally compliant products and services. This ambitious goal required us to bridge several disciplines, in particular policy and law making as well as engineering, implementation and provision of services that process data.

While working on the report, the authors discovered a wide set of divergent or even incompatible notions and definitions in the relevant disciplines, e.g. sometimes privacy is used in the narrow interpretation of data confidentiality, while others see a wider interpretation, or in some fields, the terms anonymity and pseudonymity are used interchangeably. Given the development of each field, we believe that a unification of these notions would be an impossible task; especially since the individually established terminology has to serve a variety of purposes in the respective fields. Thus we came to the conclusion that efforts need to be made to bridge the different communities, and to create a shared understanding of problems that may lead to a future shared vocabulary.

We also note that the solutions, techniques and building blocks presented in this report are of differing maturity levels. Moreover, not all privacy issues can be tackled by a method or technology alone or at all. We will briefly discuss limits of privacy by design (cf. Section 5.1). We have to bear in mind that some issues are still open, i.e. there is no technology or method to effectively cope with them. Lastly, we will provide a set of recommendations that exhort communities and decision making bodies to improve the situation (cf. Section 5.2).

## 5.1 Limits of privacy by design

Privacy by design is a technical approach to a social problem [112]. Obviously technology cannot help with all related aspects. Especially in the field of privacy, which touches various basic rights topics, such as freedom of expression and press, or protection from discrimination, issues have to be tackled in a grander scheme by society as a whole. There is a caveat to this: a significant part of the low-level privacy invasion is the direct result of the internal functioning of technical systems. Thus, while the incentives and will to invade privacy may be social problems, the actual ability to do so is a technical problem in many instances. Thus, dealing with it at the technology level is necessary.

Apart from this general discussion, there are limitations in the details. In this section we briefly discuss some of the predominant limitations. It needs to be stressed that this list is not exhaustive. Furthermore, most of the issues discussed here are not inherent limitations and obstacles to the application of privacy by design, but limitations that all parties should be aware of and calling for more research and actions from all stakeholders.

### Fragility of privacy properties

Many privacy properties are fragile with respect to composition, i.e., if a system that fulfils a certain property is embedded within or connected to another system, it is hard to assess if that privacy property is preserved. Similarly, even if two systems that fulfil a given property are combined, the resulting composed systems might not fulfil the property that was individually provided by both systems. This non-composability can easily be proven; however, the proof is by counter-example: For instance, given a message that is considered privacy-relevant data, i.e. it should not be disclosed, and two systems, where system 1 encrypts the message and sends the cypher text over a public channel and system 2 encrypts the message and sends the key over a public channel. Both systems alone do preserve the confidentiality of the message; however, if combined in a way that the same key is used for the same message in both systems, the confidentiality of the message cannot be guaranteed. By the nature of such a proof by counter-example, it does not provide properties for composition. Here more work on privacy-preserving composition is needed. Hence in practice, it is not sufficient to argue components of a privacy system are compliant, but one must also evaluate how they have been put together to ensure the resulting system is also compliant.

### Privacy metrics and utility limitations

In general a privacy metric allows comparing two systems with the same or similar functionality with respect to a set of privacy properties. However, at the moment of writing, the authors are not aware of a general and intuitive metric; known metrics are mostly used in the context of attacks on known systems (indicating how hard a given attack on a certain system is). From an information-theoretic point of view these metrics could be generalised. Since privacy has a social component, i.e., in different societies the value of a certain data item differs, the information-theoretic point of view would fall too short for data minimisation. Hence the notion of what usage of data is minimal differs. This lack of metric limits the practicality of the data minimisation principle, since it is not clear how to construct the objective function that is to be minimised. For instance, systems claiming data minimisation properties may indeed abstain from a few data attributes, but at the same time be far from an optimal solution. It may even be the case that those systems with some first steps towards data minimisation would need an entire re-design for optimising this privacy property because there would not be a transition possibility in the chosen system construction towards further data minimisation. Moreover, the lack of a universal privacy metric makes risk assessment harder; it is hard to assign probability and economic harm to a certain privacy breach.

Moreover, deploying privacy by design methods might limit the utility of the resulting system. Hence the designer needs to find a trade-off between privacy and utility w.r.t. a certain metric For example, two approaches to construct privacy-friendly statistical databases can be observed.

(1) Prior privacy. This corresponds to privacy by design. A privacy model (k-anonymity, differential privacy, etc.) and a parameterisation are used on the original data to obtain "a priori" privacy guarantees for the resulting anonymised data. The problem here is that the resulting anonymised data may have little analytical utility.

(2) Posterior privacy. An anonymisation method with certain utility preservation features is used on the original data to make sure that the anonymised data will at least preserve certain characteristics from the original data (e.g. means and variances). After that, the extant privacy (e.g. disclosure risk) is measured. If the disclosure risk is too high, then the anonymisation method is re-run with more stringent parameters (or is replaced by another method).

It turns out that the vast majority of organisations publishing anonymised data (national statistical offices, etc.) use the posterior privacy approach. The prior privacy approach (privacy by design) is quite popular among the computer science academic community but, with the exception of k-anonymity, is seldom used in real data releases. Anonymisation is a key challenge for the next decade because this tension between privacy and utility will be at the core of the development of the big data business.

### Increased complexity

Improving privacy might add user burden and friction to a system. Mental models and metaphors from the brick and mortar world might not always work in the cyberworld. Furthermore, the complexity of the system increases even if naive implementations are considered. Privacy properties might add even more complexity, often by distribution of knowledge and power. However, this makes it more difficult to determine responsibilities if something went wrong. In addition, it is for the user not always clear why privacy by design and default matters, if consent was asked. The argument by a considerable proportion of the society is: "if I do not want to reveal a certain fact to the general public, I might better tell nobody." Even though this argument is not valid, it has prominent advocates, e.g., Eric Schmidt has said "If you have something that you don't want anyone to know, maybe you shouldn't be doing it in the first place".

### Implementation obstacles

System manufacturers and standardisation bodies are usually not addressed by the data protection law. They are both rather industry and thereby market driven. However, it seems that privacy as a product is not a success. Due to network effects, it is even to observe that privacy-intrusive products are more successful than diligently developed products in line with privacy principles. But even when privacy by design principles are applied, this does not guarantee lawfulness; for instance, the developed systems may be in conflict with other legal obligations, e.g. demanding or retaining data that would not be necessary for the purpose. Furthermore, there are no or little incentives for industry to apply privacy by design since there are no or little penalties for developments not compliant with privacy by design. Moreover, there is still a lack of design methodologies and tools that are integrated in software development environments.

### Unclear or too narrow interpretation of privacy by design

A further limitation of privacy by design occurs when it is very narrowly interpreted. Many decision makers use the terms privacy and data protection interchangeable ignoring certain aspects of privacy.

From this reductionist view on privacy stems also the strong focus on data minimisation. Even though it is true that what is not collected cannot be lost (or disclosed), for many interesting applications controlled disclosure of information is more important because some personal data has to be disclosed and the subject keeps all his legal rights on his personal data after disclosure. For example, the data has to be used only for the declared purpose, deleted when it is no longer necessary, the subject is entitled to get access to his data, etc. However, it is technologically very hard to keep control over information once it has been disclosed, hence the need for transparency and accountability.

## 5.2  Recommendations

In today's information and communication technology landscape, privacy by design usually does not happen by itself, but it needs to be promoted. There is a deficiency in awareness and knowledge among system developers and service providers. Traditional and widespread engineering approaches simply ignore privacy and data protection features when realising the desired functionality. The situation is further aggravated by a deficit of current developer tools and frameworks, which make it easy to build non-compliant systems, but nearly impossible to build a compliant one. The research community dealing with privacy and data protection is growing. However, research on privacy engineering is currently insufficiently interlinked with practice, i.e. many potential solutions go unnoticed by those who could apply or provide them, and some potential solutions are rather incompatible with the utility and functionality that is expected in a practical setting. Having said that, in several areas there is a considerable need for research on privacy and data protection issues—on the methodological-conceptual as well as the operational-practical level. Further, regulatory bodies often lack expertise as well as resources for effective enforcement of compliance with the legal privacy and data protection framework, which will explicitly demand data protection by design as soon as the European General Data Protection Regulation comes into force.

In the remainder of this section, we give recommendations for actions to be taken so that engineering privacy by design becomes reality in all sectors of development.

### Incentive and deterrence mechanisms (policy makers)

**Policy makers need to support the development of new incentive mechanisms and need to promote them**. These need to move away from a narrative of "balance" between privacy and security. Both goals are necessary and can be achieved, and in fact in many contexts they mean the same thing. Furthermore the motivation to implement privacy and data protection measures needs to move from fear-based arguments; privacy protection needs to be seen rather as an asset than as cost factor.

System developers and service providers need clear incentives to apply privacy by design methods and offer privacy-friendly and legally compliant products and services. This needs to include the establishment of audit schemes and seals taking into account privacy criteria to enable the costumer to make informed choices, but also to establish effective penalties for those who do not care or even obstruct privacy-friendly solutions. As set forth in a recent annual report by the French Council of State [211], for systems processing sensitive data or for which a high level of risk has been identified, data processors should have an obligation of periodic privacy assessments operated by independent certification bodies. Further, we recommend that when DPAs consider penalties for breaches or violations they take into account the degree to which organisations provided technical controls such as the implementation of PETs to protect privacy and data.

Public services must serve as a role model by increasing the demand of privacy by design solutions. Hence public services should be legally obliged to only use systems and services where the provider can convincingly demonstrate that those are privacy-friendly and compliant with privacy and data protection law. Privacy and data protection criteria have to become part of the regular procurement procedures. Furthermore, infrastructures should implement and support privacy and data protection functionality as much as possible. This will create a market for privacy-friendly services, which also gives a choice to individuals or companies.

Funding agencies of member states or the EU should require a successful privacy and data protection assessment of each project and planned or achieved results. This should begin before deciding on funding or subsidies. Funding may also play a role for actively promoting privacy by design where market forces have not been working in favour. For instance, all developed products and services have to compete with well-established, yet privacy-invasive business models, e.g. when services on the Internet are offered for "free" by harvesting personal data of users. Here other models have to be sought—which may involve support by state money for products or services that become essential for the information society or taxing mechanisms (for example on the processing of personal data) as already proposed in several studies. This is in particular true for infrastructures, e.g. when setting up and providing communication systems that guarantee end-to-end encryption as well as anonymity on a basic layer to protect citizens against analysis of content and metadata by service providers or secret services.

A revision of policy is necessary to eliminate loopholes. In particular, legal compliance of a product or service should not be possible if it is not designed under the privacy by design paradigm. Here special attention has to be given to data minimisation as a default, comprehensive and easily accessible information for users and support for the rights of the persons concerned. Seeking the user's consent should not allow for a disproportionate processing of personal data. Having said that, the burden of this process needs to be lowered cf. our recommendation *practical support*.

Furthermore, data protection authorities should be equipped with extended mandates and funds to be able to actively search for data protection violations. This instrument can only be effective if sufficient deterrence mechanisms are in place. However penalties are not necessarily the most deterring factor for all industries (and even less for states) and negative publicity can be a much more effective deterrent in many cases. Therefore, data protection authorities should also have the duty to make public all data protection and privacy breaches and their extent.

### Promoting a deeper understanding (R&D community, education, media, policy makers)

**The research community needs to further investigate in privacy engineering, especially with a multidisciplinary approach. This process should be supported by research funding agencies. The results of research need to be promoted by policy makers and media.** Currently the research results do not provide answers and solutions to all questions concerning privacy engineering. Here the communities from various disciplines—computer science, engineering, law, economics, psychology, sociology, ethics, to name only a few—should continue and even strengthen their efforts in research. Especially cross-disciplinary collaboration is necessary when looking for comprehensive, future-proof solutions on how to design systems and services that respect the values of our society. This should be supported by funding agencies as well as the individual research communities that traditionally focus on their own methodology and hardly encourage thinking outside the box. Cross-disciplinary work should not be harmful to a researcher's career or count less than other approaches when assessing the achievements of a scientist. Interdisciplinarity should be promoted in a more systematic and pro-active way, for example by ensuring that selection committees in calls for proposals or position offers are interdisciplinary. In the absence of such measures interdisciplinarity is bound to remain a pious wish.

Better understanding should also be increased with respect to policy makers. Several initiatives have already been taken in Europe to improve the interactions between government bodies, parliaments, academics, and civil society. For example, in June 2014 the French national assembly has set up a commission on "rights and freedom in the digital age"[63] with equal representation from members of parliament and experts and the Italian parliament has taken a similar initiative[64]. These efforts should be encouraged at the European level to ensure a better mutual understanding of the specific issues raised by information technologies, especially with respect to privacy.

Also, the discussion on operational privacy engineering is up to now a mainly a technological discussion; however, it has direct implications on the organisational structure of a service or product as well as on the business models and processes. Data processing entities, vendors and manufacturers of products and service providers should keep in mind that a holistic approach is necessary for building in privacy and data protection features. Hence privacy by design methodologies should and already do cover non-technological aspects.

In addition, education programs for raising awareness as well as transmitting knowledge need to be introduced for different target groups. This comprises a hands-on approach for software developers and system providers, tackling of policy-related considerations for legislators and decision makers and practical advice for data-processing entities looking for products and services to be used for their purposes. Particular attention should be paid to those concepts and modules that may seem counter-intuitive and are not easily translated into mental models, e.g. functionality based on zero-knowledge proofs. It would be good if teachers and media would contribute to informing, enlightening and training people to foster understanding, practical usage of tools and a competent debate. Having said that, privacy researchers need to be encouraged to increase their practical knowledge, i.e., they need to better understand the requirements of the applications for which they provide solutions. In addition it is to be observed that some proposed solutions scarify the utility of the original system, hence privacy researchers need to become more knowledgeable on the application fields for which they provide privacy-friendly technologies

While the above described focus on the practical implementation of privacy-enhancing technology is of outermost importance, we also stress that fundamental research is still needed. Funding agencies need to support basic research on unconditional privacy enhancing technologies. In the past it was observed that there was a tendency to support research only if it took law enforcement mechanisms into account from the very beginning. However, the authors believe that this inhibits scientific progress.

### Practical support (R&D community, data protection authorities)

**Providers of software development tools and the research community need to provide tools that enable the intuitive implementation of privacy properties**. Privacy engineering should become easier. At the moment we observe a lack of engineers that are knowledgeable in the fields of security and privacy. While it is a noble aim to spread the word and to educate more engineers on this topic, similar efforts need to be invested to make the fields more accessible. The following actions are recommended.

Development tools need to have integrated functionalities that make the design and implementation of privacy properties intuitive. These tools should integrate freely available and maintained components with open interfaces and APIs. They should also make it possible to consider privacy by design as a continuous process encompassing all stages of the life cycle of a product or system, from risk analysis to design, implementation, exploitation and accountability.

**Especially in public co-founded infrastructure projects, privacy supporting components, such as key servers, anonymising relays, should be included.** PETs need infrastructure support from key servers,

---

[63] http://www2.assemblee-nationale.fr/14/commissions/numerique
[64] http://www.camera.it/leg17/537?shadow_mostra=23964

to encryption libraries to anonymising relays. This support for operations lowers the cost of deploying PETs for all users and operators.

Furthermore, best and good practice guides need to be published and kept up-to-date involving the participation of the data protection authorities. These guides need to come with recommendations of best available techniques and state of the art reviews. A focus should be put on the question of the appropriate defaults for privacy and data protection. Similar to the assessment of the security level of cryptographic algorithms, independent experts should contribute to a description of features, dependencies and the maturity of techniques and tools. For clarity on how to implement the European data protection framework in general and provisions on data protection by design and by default (Art. 23 GDPR) in particular, the data protection authorities should reach a consensus on a pan-European level on the requirements stemming from the legal data protection framework and publish their evaluation results as well as joint recommendations. A starting point would be—as recommended by [174]—"identifying best practices in privacy design and development, including prohibited practices, required practices, and recommended practices". For Internet-related aspects, the Internet Privacy Engineering Network (IPEN)[65] founded by the European Data Protection Supervisor together with other data protection authorities could play a central role in this endeavour. Also open-source and free software modules and tools should be collected and maintained in a readily available repository to offer a toolbox for system developers and service providers. The maintenance of the code should be guaranteed by provisioning public grants to consortia consisting of experts from research, practice and regulatory bodies, in particular data protection authorities.

**Data protection authorities will play an important role for independently providing guidance and assessing modules and tools for privacy engineering.** As suggested in [211], they could also play a key role in the promotion of privacy enhancing technologies and the implementation of the transparency principle (in particular with respect to the use of algorithms which can have a significant impact on the life of individuals).For this, they need a clear mandate as well as appropriate resources, in particular technical staff with experience in evaluating privacy and data protection criteria. IPEN together with other data protection authorities could support this activity.

### Norms (legislators, standardisation bodies)

Legal norms and ICT-related standards as well as social norms influence our lives to a large extent. Social norms predominantly play a role in the interaction of people; they evolve over time and usually cannot be set by order of a state or an organisation. In contrast, legislators and standardisation bodies expect that their output is being adhered to.

**Legislators need to promote privacy and data protection in their norms.** For privacy and data protection by design becoming reality, it would be good if legislators supported the principles from the legal European data protection framework by harmonised law as far as relations to privacy and data protection can be identified. This comprises all kinds of legal norms that enlist personal data to be processed—here it should be checked whether all the single data items are really necessary for the purposes—as well as norms containing retention periods. Obviously national and European norms on cybersecurity or on personal identification have a relation, but even tax law or procurement law are affected and should be adjusted as much as possible. It would not be sufficient to refer to general data protection rules or to shallowly mention "privacy by design" if this is not rooted into the design of the respective law. Even if it is not possible to create and maintain an apparatus of consistent legal norms, improvements are needed so that solutions that advance the state of the art of privacy and data protection are not—inadvertently or deliberately—excluded because they manage with less personal data than the legacy systems or need some initial investment.

---

[65] https://secure.edps.europa.eu/EDPSWEB/edps/EDPS/IPEN

**Standardisation bodies need to include privacy considerations in the standardisation process.** A similar situation occurs with ICT-related standards: consistency is hard to achieve, and in particular privacy and data protection requirements are often seen as additional demands from local law that will not be part of internationally agreed standards. It would be helpful for promoting privacy and data protection by design if those standards that come into effect for Europe explicitly embrace privacy and data protection instead ignoring it or—what is even worse—producing standards that may obstruct better privacy. So standardisation bodies and policy makers should integrate privacy by design in their ICT-related standards and policies. For this process it might be necessary to re-evaluate the advantages and disadvantages to keep policy and legislation technology neutral.

**Standardisation bodies need to provide standards for privacy features that ensure interoperability.** For several privacy features standardisation would be helpful. For instance, interfaces between modules could be standardised to make sure that solutions with built-in privacy properties are not excluded. This would also promote the possibility of integrating privacy solutions. In the field of transparency, standardised policy languages that express descriptions of data processing and safeguards as well as notifications of risks or how to exercise user rights would raise awareness and comprehension of individuals to another level. In particular, standardised ways for communication of privacy properties to the user (in a way interpreted by human beings, e.g. via privacy icons)[66] as well as to the user's device (in a machine-readable way) would support transparency and—if designed to support the individual's choices and interventions—also user control and self-determination. Standardisation in this area would pave the way for users to compare different products and services with respect to their privacy guarantees, and it would also make compliance checks easier for data protection authorities. In addition, it could favour the development of collaborative platforms in which users can take part in the elaboration of the privacy policies [211] (rather than being trapped into a "take it or leave it" situation). Standardisation on a European level would have to be in line with the legal European data protection framework, and it could contribute to a world-wide understanding—or even a global consensus—on privacy and data protection.

---

[66] One first example of successful standardisation by the European Commission is the EU-wide logo for Radio Frequency Identification (RFID) smart chips and systems (European Commission: Digital privacy: EU-wide logo and "data protection impact assessments" aim to boost the use of RFID systems. Brussels, 30 July 2014. http://europa.eu/rapid/press-release_IP-14-889_en.htm). In addition to the standardised signet, a privacy and data protection impact assessment process has been defined that will be reviewed by national data protection authorities prior to the use of the RFID chips.

## Annex A:    The policy context

Recently the EU published policy documents that promote 'by design' and 'by default' principles for security, privacy, and data protection after having expressed the demand for promoting privacy-enhancing technologies already in 2007 [67]. Below we exemplarily list some of these policy documents, relevant for the EU perspective, providing the context and the expectations for these measures. This section extends the brief description in Sections 2.1 and 2.2 to show the trends in the European policy context towards privacy by design.

### A.1   Strategic policy documents related to ICT and cyberspace

Digital Agenda for Europe. The Digital Agenda for Europe[67], one of the European Commission (EC) initiatives of the Europe 2020 Strategy, identifies policies and actions to maximise the benefits of Information and Communication Technologies. Actions are proposed as part of the modernisation of the European personal data protection regulatory framework "in order to make it more coherent and legally certain". For example, action #4 is specifically dedicated to the "review of the European data protection regulatory framework with a view to enhancing individuals' confidence and strengthening their rights". In the Digital Agenda, it is acknowledged that "The right to privacy and to the protection of personal data are fundamental rights in the EU which must be—also online—effectively enforced using the widest range of means: from the wide application of the principle of 'Privacy by Design' [...] in the relevant ICT technologies, to dissuasive sanctions wherever necessary".

The Cybersecurity strategy of the European Union. As a step to implement the Digital Agenda, the European Commission, the High Representative of the Union for Foreign Affairs and Security Policy, have published a cybersecurity strategy for the European Union[68] in February 2013. The cybersecurity strategy "An Open, Safe and Secure Cyberspace"[69] provides a list of priorities and actions aimed at enhancing cyber resilience of information systems, reducing cybercrime and strengthening EU international cybersecurity policy and cyber defence, while promoting values of freedom and democracy and ensuring the safe growth of digital economy. As a part of this strategy, relevant stakeholders are invited to "[s]timulate the development and adoption of industry-led security standards, technical norms and security-by-design and privacy-by-design principles by ICT product manufacturers and service providers, including cloud providers; new generations of software and hardware should be equipped with stronger, embedded and user-friendly security features". All these objectives are part of the priorities and actions to promote a single market for cybersecurity products.

### A.2   Personal data protection policy documents in the EU

Already the ePrivacy Directive[70] and especially Art. 14.3 can be interpreted as call for privacy by design; it empowers the Commission to set forth rules how to design terminal equipment in such "*a way that is compatible with the right of users to protect and control the use of their personal data, in accordance with Directive 1999/5/EC and Council Decision 87/95/EEC of 22 December 1986 on standardisation in the field of information technology and communications*".

---

Several documents of Art. 29 Data Protection Working Party ask for and refer to privacy by design. For instance in the working paper 168 on the Future of Privacy [23], it is required to "Innovate the framework [data protection framework, editors' comment] by introducing additional principles (such as 'privacy by design' and 'accountability')." Furthermore, the principle of privacy by design is explained and reasoning for this principle is provided.

> *"Users of ICT services—business, public sector and certainly individuals—are not in a position to take relevant security measures by themselves in order to protect their own or other persons' personal data. Therefore, these services and technologies should be designed with privacy by default settings.*

> *46. It is for these reasons that the new legal framework has to include a provision translating the currently punctual requirements into a broader and consistent principle of privacy by design. This principle should be binding for technology designers and producers as well as for data controllers who have to decide on the acquisition and use of ICT. They should be obliged to take technological data protection into account already at the planning stage of information-technological procedures and systems. Providers of such systems or services as well as controllers should demonstrate that they have taken all measures required to comply with these requirements.*

> *47. Such principle should call for the implementation of data protection in ICT (privacy by design or 'PbD') designated or used for the processing of personal data. It should convey the requirement that ICT should not only maintain security but also should be designed and constructed in a way to avoid or minimize the amount of personal data processed. This is in line with recent case law in Germany.*

> *48. The application of such principle would emphasize the need to implement privacy enhancing technologies (PETs), 'privacy by default' settings and the necessary tools to enable users to better protect their personal data (e.g., access controls, encryption). It should be a crucial requirement for products and services provided to third parties and individual customers (e.g. WiFi-Routers, social networks and search engines). In turn, it would give DPAs more powers to enforce the effective implementation of such measures."*

Practical aspects are also highlighted:

> *"In practice, the implementation of the privacy by design principle will require the evaluation of several, concrete aspects or objectives. In particular, when making decisions about the design of a processing system, its acquisition and the running of such a system the following general aspects / objectives should be respected:*

> - *Data Minimization: data processing systems are to be designed and selected in accordance with the aim of collecting, processing or using no personal data at all or as few personal data as possible.*

> - *Controllability:[71] an IT system should provide the data subjects with effective means of control concerning their personal data. The possibilities regarding consent and objection should be supported by technological means.*

> - *Transparency: both developers and operators of IT systems have to ensure that the data subjects are sufficiently informed about the means of operation of the systems. Electronic access / information should be enabled.*

---

[71] Throughout the report we used intervenability for this aspect.

- *User Friendly Systems: privacy related functions and facilities should be user friendly, i.e. they should provide sufficient help and simple interfaces to be used also by less experienced users.*

- *Data Confidentiality: it is necessary to design and secure IT systems in a way that only authorised entities have access to personal data.*

- *Data Quality: data controllers have to support data quality by technical means. Relevant data should be accessible if needed for lawful purposes.*

- *Use Limitation:[72] IT systems which can be used for different purposes or are run in a multi-user environment (i.e. virtually connected systems, such as data warehouses, cloud computing, digital identifiers) have to guarantee that data and processes serving different tasks or purposes can be segregated from each other in a secure way."*

The limitations of the privacy by design are also acknowledged.

*"The privacy by design principle may not be sufficient to ensure, in all cases, that the appropriate technological data protection principles are properly included in ICT. There may be cases where a more concrete 'hands on approach' may be necessary. To facilitate the adoption of such measures, a new legal framework should include a provision enabling the adoption of specific regulations for a specific technological context which require embedding the privacy principles in such context."*

The proposed regulation on data protection actually includes reference to data protection by design.

**The proposed regulation on data protection**. In January 2012 the European Commission proposed a regulation on data protection [159] that will replace the existing Data Protection Directive [82]. The proposal for the new regulation in general associates the requirements for data protection by design and data protection by default with data security and contains specific provisions relevant to privacy by design and by default in Article 23[73]. Article 23 "Data protection by design and by default" sets the obligations arising from these new principles, as they are referred in the proposed regulation:

*"[...] Having regard to the state of the art and the cost of implementation, the controller shall, both at the time of the determination of the means for processing and at the time of the processing itself, implement appropriate technical and organisational measures and procedures in such a way that the processing will meet the requirements of this Regulation and ensure the protection of the rights of the data subject. [...] The controller shall implement mechanisms for ensuring that, by default, only those personal data are processed which are necessary for each specific purpose of the processing and are especially not collected or retained beyond the minimum necessary for those purposes, both in terms of the amount of the data and the time of their storage. In particular, those mechanisms shall ensure that by default personal data are not made accessible to an indefinite number of individuals."*

---

[72] Throughout the report, we used purpose limitation for this aspect.
[73]http://eur-lex.europa.eu/LexUriServ/LexUriServ.do?uri=COM:2012:0011:FIN:EN:PDF 71

The European Parliament (EP) provided amendments[74] in January 2013; while the vote took place in March 2014[75]. The amendments voted by the European Parliament provide more explanations regarding the meaning of data protection by design and data protection by default. For instance, amendment 33 rephrases preamble (61) and clarifies: *"The principle of data protection by design requires data protection to be embedded within the entire life cycle of the technology, from the very early design stage, right through to its ultimate deployment, use and final disposal. This should also include the responsibility for the products and services used by the controller or processor. The principle of data protection by default requires privacy settings on services and products which should by default comply with the general principles of data protection, such as data minimisation and purpose limitation."*

Also, Article 23 is extended, and the voted version is of alignment 1 of Article 23 is

> *"Having regard to the state of the art, current technical knowledge, international best practices and the risks represented by the data processing, the controller and the processor, if any, shall, both at the time of the determination of the purposes and means for processing and at the time of the processing itself, implement appropriate and proportionate technical and organisational measures and procedures in such a way that the processing will meet the requirements of this Regulation and ensure the protection of the rights of the data subject, in particular with regard to the principles laid out in Article 5. Data protection by design shall have particular regard to the entire lifecycle management of personal data from collection to processing to deletion, systematically focusing on comprehensive procedural safeguards regarding the accuracy, confidentiality, integrity, physical security and deletion of personal data. Where the controller has carried out a data protection impact assessment pursuant to Article 33, the results shall be taken into account when developing those measures and procedures."*

Furthermore, according to Article 30, the future European Data Protection Board shall be entrusted with the task of issuing guidelines, recommendations and best practices taking account of developments in technology and solutions for privacy by design and data protection by default.

While in both versions there are references to principles such privacy by design and data protection by design, in this document we assume they refer to same principle and we are not going to create a separate analysis.

## A.3  Other international conventions relevant for Europe

**OECD guidelines governing the Protection of Privacy and Transborder Flows of Personal Data.** In 1980, the OECD adopted the Guidelines Governing the Protection of Privacy and Transborder Flows of Personal Data ("1980 Guidelines") to address concerns arising from the increased use of personal data and the risk to global economies resulting from restrictions to the flow of information across borders [151]. The 1980 Guidelines, which contained the first internationally agreed-upon set of privacy principles, have influenced legislation and policy in OECD Member countries and beyond.

---

[74] European Parliament report on the Data Protection Regulation: http://www.europarl.europa.eu/meetdocs/2009_2014/documents/libe/pr/922/922387/922387en.pdf European Parliament report on the Data Protection Directive: http://www.europarl.europa.eu/meetdocs/2009_2014/documents/libe/pr/923/923072/923072en.pdf EC memo "Commission welcomes European Parliament rapporteurs' support for strong EU data protection rules", 8th on January 2013, available at: http://ec.europa.eu/commission_2010-2014/reding/pdf/m13_4_en.pdf

[75] 6\EP amended version of March 2014: http://www.europarl.europa.eu/sides/getDoc.do?pubRef=-//EP//TEXT+ [96]

The OECD Guidelines have been as amended on 11 July 2013 [152]. The revised version, in its Supplementary explanatory memorandum, refers to privacy by design as a practical implementation concepts where *"technologies, processes, and practices to protect privacy are built into system architectures, rather than added on later as an afterthought"*.

**Council of Europe Convention for the Protection of Individuals with regard to Automatic Processing of Personal Data.** The Convention for the Protection of Individuals with regard to Automatic Processing of Personal Data, drawn up within the Council of Europe was opened for signature by the member States of the Council of Europe on 28 January 1981 in Strasbourg.

In 2012, a modernisation proposal has been adopted, and in this proposal the principle of privacy by design is included in Article 8bis – Additional obligations:

> "[...] design data processing operations in such a way as to prevent or at least minimise the risk of interference with those rights and fundamental freedoms.
>
> [...] Each Party shall provide that the products and services intended for the data processing shall take into account the implications of the right to the protection of personal data from the stage of their design and facilitate the compliance of the processing with the applicable law."[76]

## A.4   Summary

As can be seen from previous sections, several policy documents refer to privacy/data protection by design and by default either as principles or as practical implementation concepts.

While these terms are not clearly defined in all these reviewed documents, they have a common focus, namely, to incorporate all privacy/data protection principles through all the design and use stages of data processing and this to be the norm by default.

The policy context section focuses only on the EU perspective; where, as seen in Section A.2 a data protection reform is ongoing. Furthermore, there are more documents not mention here that influence policy and may have to be considered in system design, e.g. resolutions of the European Privacy and Data Protection Commissioners' Conference or the International Conference of Data Protection and Privacy Commissioners, opinions and working papers of the Art. 29 Working Party or court rulings (on the European level: European Court of Human Rights, European Court of Justice).

Besides the policy documents, there are also standardisation initiatives supporting the area, e.g. from ISO/IEC [125].

---

[76] CoE, T-PD_2012_04_rev4_E, revision of ETS No. 108, available at: http://www.coe.int/t/dghl/standardsetting/dataprotection/TPD_documents/T-PD(2012)04Rev4_E_Conve

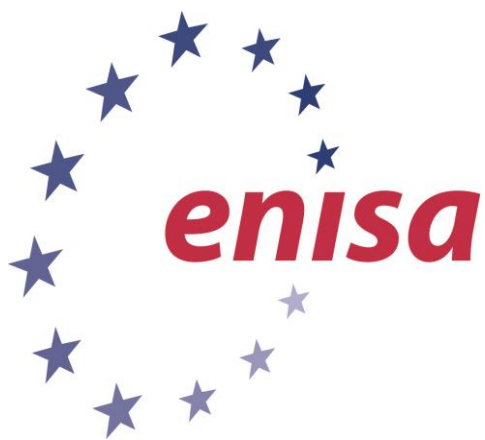

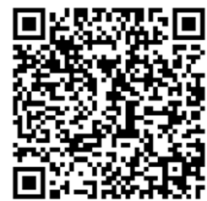

TP-05-14-111-EN-N

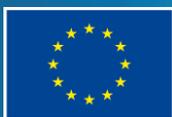

PO Box 1309, 710 01 Heraklion, Greece
Tel: +30 28 14 40 9710
info@enisa.europa.eu
www.enisa.europa.eu